# Shape-Dependence of Spontaneous Photon Emission by Quantum Electron Wavepackets and the QED Origin of Bunched Electron Beam Superradiance

Bin Zhang[1], Reuven Ianconescu[1,2], Aharon Friedman[3], Jacob Scheuer[1], Mikhail Tokman[3], Yiming Pan[4,†], Avraham Gover[1,†]

[†]Corresponding authors: gover@eng.tau.ac.il; yiming.pan@shanghaitech.edu.cn

[1] School of Electrical Engineering - Physical Electronics, Center of Light-Matter Interaction, Tel Aviv University, Ramat Aviv 69978, Israel

[2] Shenkar College of Engineering and Design 12, Anna Frank St., Ramat Gan, Israel

[3] Schlesinger Family Accelerator Centre, Ariel University, Ariel 40700, Israel

[4] School of Physical Science and Technology and Center for Transformative Science, ShanghaiTech University, Shanghai 200031, China

**Abstract**

It has been shown that the spontaneous emission rate of photons by free electrons, unlike stimulated emission, is independent of the shape or modulation of the quantum electron wavefunction (QEW). Nevertheless, here we show that the quantum state of the emitted photons is non-classical and does depend on the QEW shape. This non-classicality originates from the shape dependent off-diagonal terms of the photon density matrix. This is manifested in the Wigner distribution function and would be observable experimentally through Homodyne detection techniques as a squeezing effect. Considering a scheme of electrons interaction with a single microcavity mode, we present a QED formulation of spontaneous emission by multiple modulated QEWs through a build-up process. Our findings indicate that in the case of a density modulated QEWs beam, the phase of the off-diagonal terms of the photon state emitted by the modulated QEWs is the harbinger of bunched beam superradiance, where the spontaneous emission is proportional to $N_e^2$. This observation offers a potential for enhancement of other quantum electron interactions with quantum systems by a modulated QEWs beam carrying coherence and quantum properties of the modulation.

# Introduction

Laser-modulated free-electron beams have emerged as a promising technique for quantum states manipulation, sensing and excitation (*1–4*), leading to a new research field called "quantum electron optics". Recent advancements in high-quality coherent electron source technology have enabled the shaping and modulation of free electron wavefunction (*1–8*) using a coherent laser beam in ultrafast electron microscope and diffraction setup. Utilizing the scheme of photo-induced near-field electron microscopy (PINEM) (*9*), a single quantum electron wavepacket (QEW) can create a comb-like energy spectrum through a multiphoton emission/absorption process in the proximity of a nanostructure (*6*), a foil (*10*) or a laser-beat (*11*, *12*) (see Fig. 1). The energy modulation of the QEW subsequently converts through free-space drift into tight periodic density modulation with attosecond-level micro-bunches (*7*, *8*).

The interaction between free electrons and quantized electromagnetic field has gained significant recent interest. Theoretical studies have predicted the occurrence of single-photon cathodoluminescence into a waveguide (*13*), while recent experiments have investigated the electron-induced excitations of whispering gallery modes (*14–18*) and optical fibers (*19*). Efforts are underway to establish technologies that bridge electron microscopy and photonics (*20*, *21*) enabling atomic-scale imaging combined with quantum energy spectroscopy within an electron microscope, as well as tailoring of the spectral, spatial and quantum properties of electron-driven radiation (*21*, *22*). Particularly, there is interest in the generation of non-classical light in an on-chip microcavity of small volume and high Q-factor, which can provide a stable environment for coupling out and manipulation of the emitted photons.

The notion of shape dependence of quantum electron wave interaction of a modulated QEWs beam with a quantum system (*23–25*) has been under debate until recently. While it has been established theoretically (*26*, *27*) and experimentally (*6*, *8*, *9*, *28*) that stimulated interaction of a laser beam with a QEW is shape dependent, there has been a debate on the question whether the spontaneous emission by a QEW is also shape (or modulation) dependent (*29*, *30*). A detailed QED analysis negates any possible dependence of spontaneous photon emission on the shape or density modulation of a single electron wavefunction (*31*) (contrary to semiclassical theory (*32*)). Experimental studies have also confirmed this shape independence of spontaneous emission (*30*). Nevertheless, it has been argued that QEW modulation-dependence would emerge in the case of spontaneous superradiant emission by multiple modulation-correlated QEWs, that is, in the case that the QEWs undergo modulation through the same coherent laser beam in a PINEM process that establishes phase correlation of the density-bunched QEWs. Such a process can be seen as Dicke's spontaneous superradiance (*33*) in the quantum limit of classical coherent spontaneous emission by a density modulated point particle electron beam (*34*), which exhibits an emission rate proportional to $N_e^2$ where $N_e$ is the number of interacting electrons. A similar quadratic scaling has been claimed in "free-electron bound-electron resonant interaction" (FEBERI) effect, where multiple modulation-correlated QEWs interact with a two-level quantum system, set initially at its ground level. Early controversy regarding this multiple correlated QEWs interaction effects (*35–37*), has been nearly resolved in a comprehensive quantum analysis (*38–40*).

However, settling the fact that spontaneous emission of individual QEWs is shape (modulation) independent with the fact that multiple such QEWs can interact collectively to produce enhanced radiation, has remained elusive. In this paper, we demonstrate the superradiant excitation of a cavity single mode through the coupling of a modulation-correlated electron beam with the evanescent field of a high-Q optical micro-resonator (see Fig. 1). We point out the crucial role played by the complex (phase-dependent) bunching factor (also known as the coherence factor) of the QEWs in the establishment of the correlation between the photon emissions of the modulated QEWs. By providing a detailed phase-space presentation of the photon emission state of each single modulated QEW, we unveil the underlying mechanism that facilitates multielectron quantum correlation and demonstrate it through a Wigner Distribution (WD) presentation. We then apply an iterative method to solve the multiple-electron radiation quantum state build-up commencing with the vacuum quantum state of the cavity. We reveal that the build-up of the radiation field exhibits a growth of $N_e^2$ even when the electrons are injected randomly, as long as their QEWs are all pre-modulated by the same laser beam. In the QED formulation of the superradiance process, this counter-intuitive feature is facilitated by the automatic phase matching of the QEW bunching factors of the modulated electrons and the consideration of the off-diagonal elements of the density matrix of the built-up radiation quantum state. Additionally, we further report the features of the nonclassical radiation from single and multiple QEWs, which deviate from classical (point-particles) photon emissions. In conclusion we consider the possibility of measuring the WD of the modulated QEWs superradiance state in the photon quadrature phase-space presentation and evaluating its quantum features, such as phase-space pattern variance and squeezing effects (*41*).

# Results

## Setup and Modelling

We investigate the resonant interaction of a cavity single mode of a ring resonator with multiple quantum electron wavepackets (QEWs) (shown in fig. 1). The electron wavefunctions, momentum-modulated by a CW laser, develop into a density-bunched beam through a dispersive free drift section (*8, 32*) before interacting with the ring resonator (*15, 21, 42*). The phase-locked CW laser would establish correlation between the spatially separated electrons, leading to superradiant excitation of a photonic state inside the cavity, and expectedly, the average photon number would be quadratically dependent on the input electron number similarly to the FEBRI process (*38*). For simplicity, here we consider interaction only with a single mode of the cavity.

The analysis is based on the master equation for the joint quantum wavefunction of a single free electron and the quantized cavity mode,

$$\frac{\partial}{\partial t}\hat{\rho}_{ep}(t) = -\frac{i}{\hbar}\left[\hat{H}, \hat{\rho}_{ep}(t)\right] \qquad 1$$

where $\hat{H} = \hat{H}_0 + \hat{H}_I$ includes the free electron and photon Hamiltonians $\hat{H}_0$, and their interaction Hamiltonian $\hat{H}_I$, and $\hat{\rho}_{ep}(t)$ represents the free electron and photon joint state, which is initially disentangled (at t=0) and expressed as $\hat{\rho}_{ep}(0) = \hat{\rho}_e \otimes \hat{\rho}_{ph}$. We define the joint electron-photon basis as $|k, n\rangle = |k\rangle \otimes |n\rangle$, where $|k\rangle$ is the continuous wavevector basis ($k = p/\hbar$) and $|n\rangle$ is the Fock basis of a single cavity mode. Thus, the general form of the initial free electron wavefunction and the density matrix of the photon state in the Schrödinger picture are:

$$|\psi_e\rangle = \int dk\, c_k |k\rangle$$
$$\hat{\rho}_{ph}^i = \sum_{n,n'} \rho_{ph}^i(n, n') |n\rangle\langle n'| \qquad 2$$

where $c_k$ is the electron wavefunction in the wavevector basis and $\rho_{ph}^i(n, n') = \langle n|\hat{\rho}_{ph}^i|n'\rangle$ is the photon state density matrix in the Fock basis. For a QEW modeled as a Gaussian envelope wavefunction, the electron wavefunction after PINEM energy modulation and subsequent drift time $t_d$ ($t_d = L_d/v_0$ – see Fig. 1), is (*8, 38*):

$$c_k = \frac{1}{(2\pi\sigma_k^2)^{\frac{1}{4}}} \sum_{n=-\infty}^{+\infty} J_n(2g_L) \exp\left[-\frac{(k - k_0 - n\delta k)^2}{4\sigma_k^2} - in\phi_0\right] e^{-iE_k t_d/\hbar}, \qquad 3$$

where $g_L$ is the laser modulation coupling coefficient, $\sigma_k = \sigma_p/\hbar$ is the momentum spread of the unmodulated QEW, $\delta k = \frac{\omega_L}{v_0}$ is the quantum recoil of the QEW induced by the modulating laser beam and $\omega_L$ is the laser frequency. $\phi_0$ is the laser phase, which is important for establishing the correlation between QEWs.

According to the basis $|k,n\rangle$, we define the free electron Hamiltonian $\widehat{H}_e$ and the free photon Hamiltonian $\widehat{H}_{ph}$ to be diagonal in the wavevector basis $|k\rangle$ and Fock basis $|n\rangle$, respectively. Thus, the free Hamiltonian $\widehat{H}_0$ is

$$\widehat{H}_0 = \widehat{H}_e \otimes \hat{I}_{ph} + \hat{I}_e \otimes \widehat{H}_{ph}, \qquad 4$$

where $\widehat{H}_e|k\rangle = E_k|k\rangle$ is free electron Hamiltonian and $\widehat{H}_{ph}|n\rangle = n\hbar\omega_c|n\rangle$ is non-interacting cavity photon Hamiltonian, $\hat{I}_{ph}$ and $\hat{I}_e$ are the identity operators introduced to expand the total Hilbert spaces, respectively. The energy dispersion of a relativistic free electron is given to second order by $E_k = E_0 + v_0\hbar(\hat{k} - k_0) + \frac{\hbar^2}{2\gamma^3 m}(\hat{k} - k_0)^2$ (26). The interaction between the photonic cavity mode and the free electron can be written in a general form,

$$\widehat{H}_I = -\frac{e}{2\gamma m_e}\left(\hat{p} \cdot \hat{A}(z) + \hat{A}(z) \cdot \hat{p}\right), \qquad 5$$

where the vector potential of the mode has the form $\hat{A}(z) = A_{0,eff} f(z)(\hat{a}e^{iq_z z} + \hat{a}^\dagger e^{-iq_z z})$ with $A_{0,eff} = \eta A_0$, $\eta < 1$ where $A_0 = \sqrt{\frac{\hbar}{2\varepsilon_{eff}\omega_c V_c}}$ is the intensity of the field of the vacuum mode ($V_c$ is the quantization volume of the cavity with effective dielectric constant $\epsilon_{eff}$) (43). The coupling field with the free electron $A_{0,eff}$, is determined by the evanescent penetration into vacuum of the transverse waveguide mode. The longitudinal modes of the cavity are defined by the cavity circumference. The phase velocity of the mode is assumed to satisfy a synchronism condition with the swift electron $v = \omega_L/q_z$. Here we choose the case of fundamental harmonic resonance where the phase matched photon mode frequency is the same as the modulating laser $\omega_c = \omega_L$ and its wavevector is $q_z \simeq n_{eff}\frac{\omega_c}{c}$, $n_{eff}$ is the effective index of refraction of the dielectric waveguide mode. The function $f(z)$ defines the spatial distribution of the quantized cavity mode.

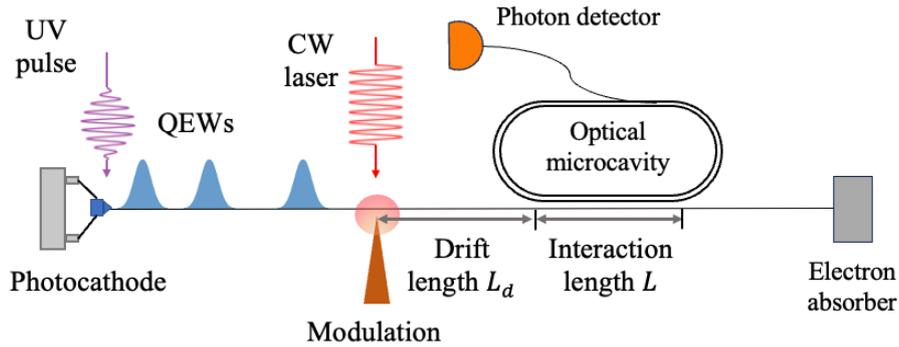

Figure 1: setup of the superradiant excitation of a cavity through a modulation-correlated QEWs beam (16, 42).

For the interaction of the cavity with one electron, the evolution of the system in the interaction picture is given by the scattering matrix $\hat{S} = \mathcal{T}\exp\left[-\frac{i}{\hbar}\int_{-\infty}^{+\infty}\widehat{H}_I^{int}(t)dt\right]$, where $\widehat{H}_I^{int}(t) = e^{i(\widehat{H}_e+\widehat{H}_{ph})t/\hbar}\widehat{H}_I e^{-i(\widehat{H}_e+\widehat{H}_{ph})t/\hbar}$. For a QEW that satisfies the large-recoil condition (26), we can expand the scattering matrix as [details in SM. 1]:

$$\hat{S} = \exp(g\hat{a}^\dagger \hat{b} - g^*\hat{a}\hat{b}^\dagger) = e^{-|g|^2/2} \sum_{n,m} \frac{1}{n!} \frac{1}{m!} (g\hat{a}^\dagger)^n (-g^*\hat{a})^m \hat{b}^{n-m}, \qquad 6$$

where the spontaneous emission coupling constant $g = i\frac{eA_{0,eff}}{\gamma m_e v_0} k_0 L$ is proportional to the vacuum electric field of the interacting radiation mode and the interaction range is $L$, the recoil operator $\hat{b} = e^{-i\frac{\omega_c \hat{z}}{v_0}}$ satisfies $\hat{b}|p_0\rangle = |p_0 - \hbar\delta k\rangle$, where $\hbar\delta k = \hbar\omega_c/v_0$ is the momentum recoil of the QEW in the PINEM process. It is worth mentioning that the photon generation operator $\hat{a}^\dagger$ exhibits the canonical commutation relation $[\hat{a}, \hat{a}^\dagger] = 1$, while the recoil operator $\hat{b}$ follows a different relation, $[\hat{b}, \hat{b}^\dagger] = 0$. The matrix elements of the scattering matrix in the combined basis are

$$\langle k_f, n_f | \hat{S} | k_i, n_i \rangle = \delta_{k_f + n_f \delta k, k_i + n_i \delta k} M_{n_f, n_i}$$
$$M_{n_f, n_i} = \langle n_f | \hat{D}(g) | n_i \rangle, \qquad 7$$

where $M_{n_f, n_i}$ is the transition matrix between Fock state $|n_f\rangle$ and $|n_i\rangle$, $\hat{D}(g)$ is the displacement operator for the photon state that serves as the generator of a coherent state $\hat{D}(g)|0\rangle = |g\rangle$, the explicit form of $M_{n_f, n_i}$ is presented in Eq. (S32). Additionally, the delta function $\delta_{k_f + n_f \delta k, k_i + n_i \delta k}$ accounts for momentum conservation of the interaction between the electron and the cavity. For the initial joint state in the general form: $\hat{\rho}_{ep}^i = |\psi_e^I\rangle\langle\psi_e^I| \otimes \hat{\rho}_{ph}^{i,in}$ ($|\psi_e^I\rangle = e^{i\tilde{H}_e t/\hbar}|\psi_e\rangle$ is the electron state and $\hat{\rho}_{ph}^{i,in}$ is the photon density matrix both in the interaction picture), the final joint state after interaction which is the solution of Eq. 1, can be obtained through: $\hat{\rho}_{ep}^f = \hat{S}\hat{\rho}_{ep}^i \hat{S}^\dagger$. Then, the final photon state starting from an arbitrary initial photon state $\rho_{ph}^{i,in}$, can be derived through tracing out the electron degrees of freedom $\rho_{ph}^{f,in} = \text{Tr}_e\{\hat{\rho}_{ep}^f\}$ (details in SM. 4):

$$\tilde{\rho}_{ph}^f(n_f, n_f + s) = \sum_{n_i} M_{n_f, n_i} \left[\sum_{r=-\infty}^{+\infty} \tilde{b}^{(-r)} \tilde{\rho}_{ph}^i(n_i, n_i + r + s)\right] M_{n_i + r + s, n_f + s}^\dagger, \qquad 8$$

where $\tilde{\rho}_{ph}^i(n_i, n_i + r + s) = \langle n_i | \hat{\rho}_{ph}^{i,in} | n_i + r + s \rangle$ and $\tilde{\rho}_{ph}^f(n_f, n_f + s) = \langle n_f | \hat{\rho}_{ph}^{f,in} | n_f + s \rangle$ are the matrix elements of the initial and final photon state, respectively. The coherence factors

$$\tilde{b}^{(n)} = \langle \psi_e^I | \hat{b}^n | \psi_e^I \rangle \qquad 9$$

is the n$^{\text{th}}$ order harmonics bunching factor of the input modulated QEW, which can be expressed as $b^{(n)} = \int dk \, c_k c_{k+n\delta k}^* e^{-in\delta k v_0 t_i}$ by substituting the electron wavefunction of Eq. 2 in the expression of $b^{(-r)}$ with $n = -r$. The contribution of the QEW to the electron-photon interaction is embodied in the bunching factors $b^{(-r)}$ in the square bracket, which represents a decoherence effect (44). Note that in the case of spontaneous emission (see next section), the initial photon state is the vacuum state $\tilde{\rho}_{ph}^i(n_i, n_i') = \delta_{n_i,0}\delta_{n_i',0}$.

The summation in the square bracket in Eq. (8) can be presented as a Hadamard product between the density matrix of the initial photon state $\rho_{ph}^{i,in}$ and a dephasing matrix $\hat{B}(s)$, namely $\mathcal{B}(s) \odot \hat{\rho}_{ph}^{j-1,in}$, where the elements of $\mathcal{B}(s)$ are defined in terms of the bunching factors of the input QEW: $\hat{B}_{nm}(s) = \tilde{b}^{(s+n-m)}$. Thus, the s-th diagonal terms $\tilde{\rho}_{ph}^f(n_f, n_f + s)$ of the final photon state Eq. 8 can be re-expressed as $\langle n_f | M \cdot [\mathcal{B}(s) \odot \hat{\rho}_{ph}^{i,in}] \cdot M^\dagger | n_f + s \rangle$. According to the definition of the dephasing matrix $\hat{B}(s)$, the s-th diagonal array of $\rho_{ph}^f$ is composed of contributions from the product of s-th diagonal array of $\tilde{\rho}_{ph}^i$ and $\tilde{b}^{(0)}$, the product of (s+1)-th diagonal array of $\tilde{\rho}_{ph}^{i,in}$ and $\tilde{b}^{(-1)}$, the product of

(s-1)-th diagonal array of $\tilde{\rho}_{ph}^{i,in}$ and $\tilde{b}^{(1)}$ and so on. This matrix multiplication process is illustrated by the first three blocks in Figure 2, starting from a photon state $\hat{\rho}_{ph}^{i,in}$ denoted as $\hat{\rho}_{ph}^{j-1,in}$.

Following the process of matrix multiplication in Figure 2, we can state that the matrix element of the final photon state (marked by x in the fourth block), is composed of contributions mainly by the elements near the same position (marked by x) in the initial photon state (region marked by a yellow square in the second block). Under perturbation theory, when the coupling is small, an element $(n, m)$ of the final photon density matrix in Eq. 8 can be written to second order of the coupling constant $g$:

$$\rho_{ph}^{(f)}(n,m) \approx e^{-g^2} \Big[ \rho_{ph}^{(i)}(n,m) + g\sqrt{m}\, b^{(-1)} \rho_{ph}^{(f)}(n, m-1) + g^*\sqrt{m+1}\, b^{(1)} \rho_{ph}^{(f)}(n, m+1)$$

$$+ g\sqrt{n}\, b^{(1)} \rho_{ph}^{(f)}(n-1, m) + g^*\sqrt{n+1}\, b^{(-1)} \rho_{ph}^{(f)}(n+1, m)$$

$$+ g^2 \sqrt{n}\sqrt{m+1}\, b^{(2)} \rho_{ph}^{(f)}(n-1, m+1) \qquad\qquad 10$$

$$+ g^2 \sqrt{n+1}\sqrt{m}\, b^{(-2)} \rho_{ph}^{(f)}(n+1, m-1)$$

$$- g^2 \sqrt{n+1}\sqrt{m+1}\, \rho_{ph}^{(f)}(n+1, m+1) - g^2 \sqrt{n}\sqrt{m}\, \rho_{ph}^{(f)}(n-1, m-1) \Big]$$

Thus, the diagonal array of the final photon density matrix $\rho_{ph}^f$ which represents the Fock state distribution, is mainly the superposition of the 1st and 2nd diagonal array terms of the dephased initial photon state $\hat{B}(0) \odot \hat{\rho}_{ph}^i$ as long as the off-diagonal matrix elements are not zero, which is dependent on the QEW shaping or modulation. The non-classical properties of the emission state are represented by the off-diagonal terms of $\rho_{ph}^f$, where the s-th diagonal array is determined by the dephased initial state: $\hat{B}(s) \odot \hat{\rho}_{ph}^i$. Hence, we concluded that the photon emission state is dependent on the QEW shaping through the bunching factors of the input electron wavepacket.

For the case of spontaneous emission, starting from the vacuum state $\tilde{\rho}_{ph}^i(n_i, n_i') = \delta_{n_i,0}\delta_{n_i',0}$, we get from Eq. 10 that the matrix element of the final photon state is:

$$\rho_{ph}^{(f)}(n,n) \approx e^{-g^2} \delta_{n,0}$$

$$\qquad\qquad 11$$

$$\rho_{ph}^{(f)}(n, n+1) = e^{-g^2} g\sqrt{m}\, b^{(-1)} \delta_{n,0}$$

This manifests that the diagonal terms of the spontaneous emission matrix are independent of the QEW bunching. However, the quantum state develops off-diagonal terms that are dependent on the bunching factors. This is most important for further interaction of the cavity photons with subsequent incoming modulated QEWs.

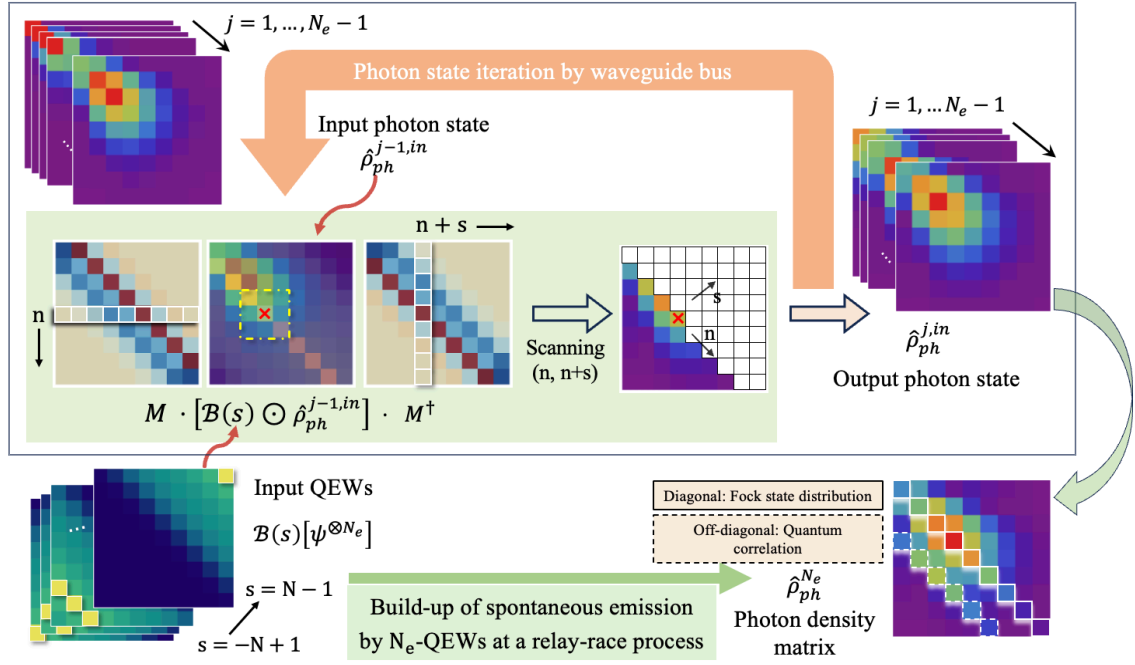

Figure 2: the iterative process of the photon-electron interaction. In j-th round of interaction, the input QEW acts as the dephasing matrix $B(s)$ located in the lower-left corner, while the input photon state $\rho_{ph}^{j-1,in}$ serves as the initial state in Eq. 8, depicted in the upper-left corner. The process of obtaining the final photon state $\rho_{ph}^{f,in}$ from the initial photon state $\rho_{ph}^{j-1,in}$ through the electron-photon interaction is visualized by the four blocks within the light green box, where the transition matrix $M$ is the matrix form of displacement operator. The resulting photon state $\rho_{ph}^{f,in}$ is then considered as the initial photon state for the (j+1)-th round interaction, denoted as $\rho_{ph}^{j,in}$. After interacting with $N_e$ electrons, the photon state within the microcavity is $\rho_{ph}^{N_e}$, shown in the lower-right corner.

The single electron interaction with a single cavity mode can now be extended to multi-electron interaction with the photons in the mode assuming a high-Q cavity, so that the photon decay can be neglected (e.g., $2ns$). Furthermore, we assume that the electron density is small enough so that there is no overlap between the QEWs[1]. However, we can iteratively apply the analysis of a single electron interaction process as long as the electron passing by the cavity is not measured. We therefore proceed with the iterative analysis of cumulative photon emission into the single cavity mode by multiple ($j = 1..N_e$) consecutive optically modulated QEWs as illustrated in Fig. 2.

For the multiple-shots single-electron beam case, the final emission state generated by $N_e$ QEWs can be obtained through applying Eq. 8 iteratively as shown in Fig. 2. The initial photon state in the cavity is denoted as $\rho_{ph}^{(0)}$, which is the vacuum state in our work, and the photon state generated by (j-1)-th electrons is denoted as $\rho_{ph}^{(j-1)}$. As shown in Fig. 2, the output photon state $\rho_{ph}^{(j-1)}$ generated by (j-1)-th input QEW will be set as the initial photon state interacting with the j-th input QEW, where the electron state is denote as $\rho_e^{(j)} = |\psi_e^{(j)}\rangle\langle\psi_e^{(j)}|$ and the diffusion matrix $\hat{B}$ is composed of $b_j^{(r_j)}$. The interaction between the j-th electron and the remained photon state $\rho_{ph}^{(j-1)}$ in the iterative process can be written as

$$\tilde{\rho}_{ph}^{j}(n_j, n_j + s_j) = \sum_{r_j=-\infty}^{+\infty} \tilde{b}_j^{(r_j)} \sum_n \tilde{\rho}_{ph}^{j-1}(n_{j-1}, n_{j-1} + r_j + s_j) M_{n_{j-1},n} M_{n+r_j+s_j,n_{j-1}+s_j}^{\dagger}, \qquad 12$$

where $n_j$, $s_j$ represent the matrix element of the photon state $\rho_{ph}^j$.

---

[1] If we consider typical parameter of a transmission electron microscope (TEM), an electron beam current of about 10 nA and electron energy 200 keV, the mean timing difference of the in-coming electrons is $\bar{t} = 1.6 \times 10^4\ fs$. This is much larger than the typical duration of a coherent QEW of about $500\ fs$ (for beam energy spread of $0.7eV$) (8, 57).

## Spontaneous emission from a single QEW

We obtain the final photon state of the spontaneous emission from a single QEW through Eq. 8, where the initial photon state is the vacuum state $\hat{\rho}_{ph}^i = |0\rangle\langle 0|$ and the transition matrix is: $M_{n_f,n_i} = \delta_{n_i,0}\langle n_f|g\rangle$, then

$$\hat{\rho}_{ph}^f = e^{-|g|^2} \sum_{n,m} \tilde{b}^{(n-m)} \frac{g^n}{\sqrt{n!}} \frac{(g^*)^m}{\sqrt{m!}} |n\rangle\langle m|. \qquad 13$$

This form of photon state is a dephased coherent state, where the decoherence here is caused by the bunching of the free electron wavepacket, and the bunching factor $\tilde{b}^{(n-m)}$ represents the decoherence degree of the $(n-m)$ off-diagonal matrix element.

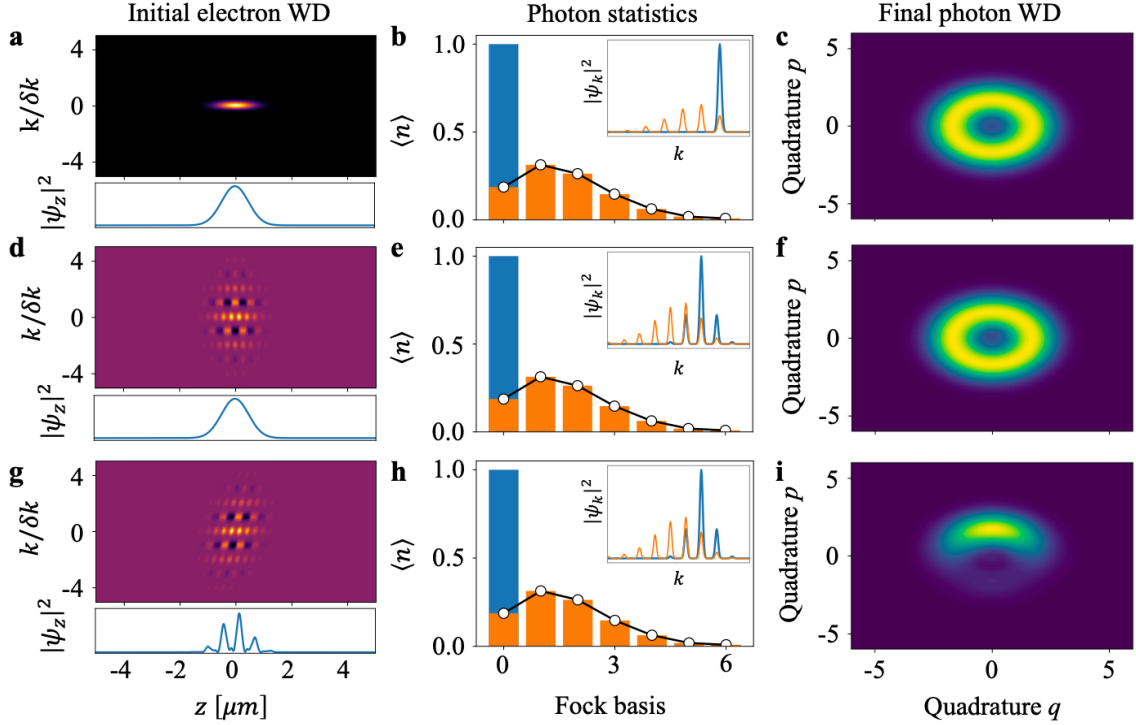

Figure 3: The final state of the cavity photon after interaction with three different QEW states with coupling constant $g = 1.3$. The initial photon state is a zero-vibration ("vacuum") quantum state (referring to spontaneous emission) and the initial QEW state is an unmodulated QEW (row 1), an energy-modulated QEW (row 2) and a density-modulated QEW (row 3). (a), (d), (g) display the free electron initial state in the phase space through Wigner distribution and in the real space through projection of its WD: we show the input electron state as (a): an unmodulated Gaussian QEW with standard deviation $\sigma_t$, (d): an energy-modulated QEW with standard deviation $\sigma_t \simeq T$ and modulation coupling constant $g_L = 0.5$, and (g): a density-modulated QEW after free drift ($t_d = 2.6 ns$) (note that the vertical projection of the tilted WD produces density modulation) . (b), (e), (h) show the final photonic states in Fock space, the pre-interaction (blue) and the post-interaction EELS (orange) are shown in the inset. The blue bar represents the initial vacuum state $|0\rangle$, and the orange bars show the diagonal terms of the final state $\rho_{ph}^{(1)}$ (Eq. 10). The black dotted line is the distribution of a coherent state with $\alpha = g$ which matches the spontaneous emission distribution derived from Eq. 10. (c), (f), (i) are the Wigner quasi-distribution functions that contain the entire information of the final photonic state Eqs. 10a and 10b, with quadrature position $\hat{q} = \frac{1}{\sqrt{2}}(\hat{a} + \hat{a}^\dagger)$ and quadrature momentum $\hat{p} = \frac{i}{\sqrt{2}}(\hat{a}^\dagger - \hat{a})$ (45).

In Figure 3, we illustrate the impact of different kinds of electron wavepackets on the spontaneous emission from the cavity, including an unmodulated QEW (a, b, c), an energy-modulated QEW (d, e, f) and a density-modulated QEW (g, h, i). As shown in the middle column, the photon statistics generated from the three types of QEWs is all the same as a coherent state distribution with an average photon number equals to $|g|^2$. This manifests the independence of the photon statistics on the shape of input QEW. This result is consistent with Eq. 13 as the photon statistics is described by the diagonal terms of its density matrix, which are all the same. The second order correlation function of the spontaneous emission states is $g^{(2)}(0) = \frac{\langle \hat{n}^2 - \hat{n} \rangle}{\langle n \rangle^2} = 1$, which is also the same as that of a coherent state $|g\rangle$.

However, the Wigner phase-space quasi-distributions of the full photon states (shown in the right column of Figure 3) demonstrate a significant distinction between the emission from a density-bunched QEW (Figure 3i) and the emission from an unmodulated or an energy-modulated QEW (Figure 3c, 3f). The emission states in 3c and 3f exhibit a ring shape, while the emission state in 3i is in a crescent shape. These differences in the emission states exhibit dependence on the shape of the QEW. They are attributed to the different distributions of bunching factors over the harmonic orders, as indicated by the general expression of the emission state Eq. 13. For the QEW state defined by Eq. 3, the bunching factors (46) can be written as (calculated under the small recoil limit, details see SM. 2):

$$\tilde{b}^{(n)} = |b^{(n)}(t_d)| e^{-in\omega_L t_d - in\phi_0}, \qquad 14$$

where the amplitude of the n$^{th}$ order bunching factor is $|b^{(n)}(t_d)| = J_n\left[4|g_L|\sin\left(n\frac{\hbar\delta k^2}{2\gamma^3 m_e}t_d\right)\right]e^{-\frac{n^2}{2}\left(\frac{\hbar t_d}{\gamma^3 m_e}\right)^2 \sigma_k^2 \delta k^2}$. This amplitude decays exponentially with the drift time from the PINEM modulation point to the microcavity - $t_d$ (see Fig. 1) and exhibits a periodicity of $T_n = T_b/n$ through the function $\sin(nt_d/T_b)$ in the argument of the Bessel function, where $T_b = 2\pi/\frac{\hbar\delta k^2}{2\gamma^3 m_e}$ is the fundamental period. For the optimal bunching position, the drift time $t_d$ is usually small, thus the decay is negligible. The case of unmodulated QEW is derived from this equation by setting the modulation coupling constant $g_L = 0$. The case of an energy-modulated QEW is derived by setting the drift time $t_d = 0$. In both cases the bunching factors of different harmonic orders satisfy $b^{(n)} = \delta_{n,0}$. Hence only the diagonal terms in Eq. 13 are non-zero. The Wigner quasi-distribution of these cases reflects this fact by showing a ring shape (Fig. 3 c, f). On the other hand, in the case of a density-modulated QEW, the emission state Eq. 13 contains non-zero off-diagonal terms because the QEW contains some harmonic orders n, make $b^{(n)} \neq 0$. This is reflected in the Wigner distribution (Fig. 3g), displaying a crescent shape.

The differences in the emission states can be explicitly demonstrated through the variance of the quadrature operators in a space, denoted as $\hat{q} = \frac{1}{\sqrt{2}}(\hat{a} + \hat{a}^\dagger)$ and $\hat{p} = \frac{i}{\sqrt{2}}(\hat{a}^\dagger - \hat{a})$. These operators satisfy the canonical commutation relation $[\hat{q}, \hat{p}] = i\hbar$. The quadrature amplitudes, which represent the amplitude fluctuations and phase fluctuations of the electromagnetic field, are two orthogonal components of the light field commonly known as the "in-phase" and "quadrature" components that can also be associated with the electric and magnetic fields of the radiation (45). The variance of x and p of the photon state given by Eq. 13 can be directly obtained by using the expression of bunching factors given in Eq. 14:

$$\Delta q^2 = \frac{1}{2} + |g|^2\left(1 - |b^{(1)}|^2\right) - |g|^2\left(|b^{(2)}| - |b^{(1)}|^2\right)\cos[2(\phi_0 + \omega_L t_d)]$$
$$\Delta p^2 = \frac{1}{2} + |g|^2\left(1 - |b^{(1)}|^2\right) + |g|^2\left(|b^{(2)}| - |b^{(1)}|^2\right)\cos[2(\phi_0 + \omega_L t_d)], \qquad 15$$

where the oscillations of $\Delta x^2$ and $\Delta p^2$ are in opposite phase around $\frac{1}{2} + |g|^2\left(1 - |b^{(1)}|^2\right)$. According to the equation above, the amplitude of the oscillation of the electric and magnetic variance is determined by the factor $|g|^2\left(|b^{(2)}| - |b^{(1)}|^2\right)$. Thus, in a QED theory, there is difference between the variances of the electric and magnetic fields of photons emitted from the shaped electron as shown in Figure 3i, signifying emission of a non-classical light. Nevertheless, while changing periodically with the drift time $t_d$, the variances of x and p are always larger than the minimal uncertainty $1/2$ since the bunching factors $b^{(n)}$ are always smaller than 1. We can define such a state as a "modified squeezed light state" (41). Notice that when $|b^{(1)}|^2 = b^{(2)}$, then $\Delta q^2 = \Delta p^2$ and the squeezing disappears.

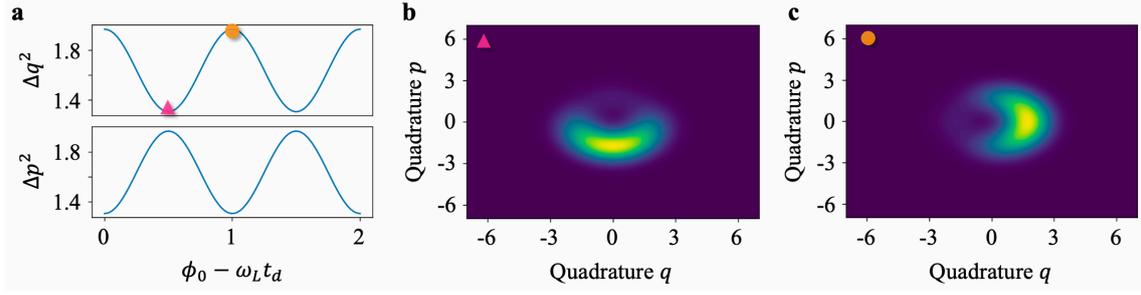

Figure 4: (a) the variance of amplitude fluctuation ($\hat{p}$) and phase fluctuation ($\hat{q}$) of the emission field from single shaped QEW for different modulation phase $\phi_0$. The oscillations of $\Delta q^2$ and $\Delta p^2$ are in opposite phase. The simulation result is computed for QEW modulation parameters: $|b^{(1)}| = 0.572$ and $|b^{(2)}| = 0.130$. The orange dashed lines at $\pi/2$ and $\pi$ correspond respectively to (b) the emission state with minimal variance of p and maximal variance of q and (c) the emission state with minimal variance of q and maximal variance of p.

According to Eq. 14 and 15, the variance of quadrature amplitudes of the EM field is determined by the phase of bunching factors, which is controlled by the modulation phase of the QEW (*47*). As shown in Figure 4b, 4c, when we tune the modulation phase $\phi_0$, the emission state rotates in phase-space. The measurement of these Wigner distributions or say the tomography of the photon state can be realized through applying a series of homodyne measurements (*48, 49*) with varying local oscillator phases. The obtained information about the probability distribution of the quadrature values, can be used to reconstruct the full photon state described by Eq. 13.

## Emission from multiple electrons

In the preceding section, we demonstrated the crucial role of the bunching factors in determining the photon state emitted by a single electron in the electron-cavity interaction. These bunching factors are determined by the shaping of the QEW. However, due to the typically small value of the coupling constant in the electron-cavity interaction ($g = 10^{-1} \sim 10^{-2}$), measuring this effect with single electrons remains challenging. In this section we show that the bunching factors of an ensemble of modulated QEWs, affect the off-diagonal elements of the density matrix of the emitted radiation quantum state, and consequently have a significant effect on the observable collective spontaneous emission of a multiple electrons beam. In particular, this enables coherent superradiant enhancement of the photon emission rate by multiple modulation-correlated QEWs [24, 25].

In the setup and modelling section, we introduced the iterative interaction process described in Eq. 12. This allows us to systematically calculate computationally the emitted photon state from $N_e$ electrons step by step. To get a more profound insight into the interaction, we derive here also an approximate analytical expression for the final emission photon state. For the typical case of electron-photon interaction, the coupling constant $g$ is small, which enables us to consider only the 1st order bunching factors of the j-th electron ($\tilde{b}_j^{(1)}$ and $\tilde{b}_j^{(-1)}$) in Eq. 12. This is especially true when the QEWs are under small modulation amplitude, since the bunching factors $\tilde{b}^{(n)}$ decay rapidly with the harmonic index n. Hence, the diagonal arrays of the density matrix of the emitted photons can be expressed in terms of a power series of the 1st order bunching factors $\tilde{b}_j$. Notice that since the density matrix is Hermitian the relation $b_j^{(-1)} \equiv b_j^{(1)*}$ is valid.

For the sake of brevity, we will henceforth drop the marking (1), as we will mostly deal with the first order bunching factor, when higher orders of the bunching factor are involved, we will keep the marking of the order.

In a superradiance scenario one must account for the accumulated effect on the photonic state of all bunched QEWs in the beam. This is especially true when that state is created in a cavity and stays there as long as incoming electrons keep contributing to it. The question then arises; how many electrons are to be taken into account? While in the numerical computational procedure we can take a large number of cooperating electrons, in the analytical derivation we derive the density matrix in terms of a combinatorial converging series of ascending number of collaborating electrons.

First, we use Eq. 12 to evaluate the 0$^{th}$ order diagonal term of the density matrix $\hat{\rho}_{ph}^{N_e}$, which represents the photon statistics. It can be expressed as

$$\rho_{ph}^{N_e}(n,n) = c_{0,n}^{(0)} + c_{0,n}^{(2)} \sum_{i \neq j}^{N_e} \tilde{b}_i \tilde{b}_j^* + c_{0,n}^{(4)} \sum_{i \neq j \neq k \neq l}^{N_e} \tilde{b}_i \tilde{b}_j^* \tilde{b}_k \tilde{b}_l^* + c_{0,n}^{(6)} \sum_{\substack{i \neq j \neq k \neq \\ l \neq m \neq q}}^{N_e} \tilde{b}_i \tilde{b}_j^* \tilde{b}_k \tilde{b}_l^* \tilde{b}_m \tilde{b}_q^* \ldots , \qquad 16$$

The coefficients $c_{0,n}^{(2m)}$ (derived in SM. 6) denote the correlated contribution to the Fock photon distribution arising from the bunching factors of any 2m correlated electrons within the $N_e$ electrons in the beam. In Eq. 16 we took into account involvement of up to six electrons at one time. It is straightforward, to extended to more electrons using combinatorial formalism. As shown in SM 6, only even numbers (2m) are included in the series where m is an integer number. Equation 16 states that the probability of observing the Fock state $|n\rangle$ is the sum of even products of $b$ coefficients involving the bunching factors of distinct electrons. Using the combinatorial relations in SM. 6 (table 1), the photon number expectation value of the emitted photon state described by the first two terms of Eq. 16 is

$$\left\langle n_{ph}^{(N_e)} \right\rangle = \sum_{n=0}^{\infty} n \rho_{ph}^{N_e}(n,n) = N_e |g|^2 + |g|^2 \sum_{i \neq j}^{N_e} \tilde{b}_i \tilde{b}_j^*. \qquad 17$$

It is noteworthy that Eq. 16 suggests an entanglement between the photonic state and the emitting electrons. The level of entanglement could be calculated from the photonic density matrix. However, it is beyond the scope of this paper.

In the absence of modulation ($\tilde{b}_j = 0$) and also when the QEWs are modulated but uncorrelated, all terms in Eq. 16 vanish (on average) except the first. In this case, the diagonal terms of the density matrix $\rho_{ph}^{N_e}(n,n)$ represent the photon statistics of spontaneous emission by unmodulated or uncorrelated QEWs, given by Eq. 16. The coefficient $c_{0,n}^{(0)}$ can be approximately expressed in an analytical formula

$$\rho_{ph}^{N_e}(n,n) = c_{0,n}^{(0)} = \frac{1}{N_e |g|^2 + 1} \left( \frac{N_e |g|^2}{N_e |g|^2 + 1} \right)^n$$

which is the Bose-Einstein distribution of super-Poissonion statistics (*50, 51*), the match between the analytical approximation and the numerical results is shown in SM. 6.

The off-diagonal terms of the density matrix $\hat{\rho}_{ph}^{N_e}$, that signify the non-classical properties of the emission, exhibit similar properties to the diagonal terms. For example, the terms in the 1$^{st}$ diagonal (super diagonal, next and above the principal diagonal) representing quantum correlation, can be expressed as (Eq. S53)

$$\tilde{\rho}_{ph}^{N_e}(n, n+1) = c_{1,n}^{(1)} \sum_{i=1}^{N_e} \tilde{b}_i^* + c_{1,n}^{(3)} \sum_{i \neq j \neq k}^{N_e} \tilde{b}_i \tilde{b}_j^* \tilde{b}_k^* + c_{1,n}^{(5)} \sum_{i \neq j \neq k \neq l \neq m}^{N_e} \tilde{b}_i \tilde{b}_j^* \tilde{b}_k \tilde{b}_l^* \tilde{b}_m^* \ldots, \qquad 18$$

where $c_{1,n}^{(2m+1)}$ denotes the contribution to the n-th component of the 1st diagonal array arising from the coupling of the bunching factors of (2m+1) electrons. Again, m is a positive integer number, so that the number of electrons in each term of Eq.18 is odd.

Hence, the final emission state is crucially dependent on the bunching factors of all the electrons involved in the interaction. Considering electrons subjected all to modulation by the same laser beam, and passing the same free propagation process as shown in Figure 1, they differ only by the modulation phase of the individual QEWs relative to the wavepacket center $\phi_0^j$ and by their random arrival time $t_j$ to the PINEM modulation point. The bunching factors of j-th input QEW, upon arrival at the interaction region, is (see supplementary (Eq. S26)):

$$\tilde{b}_j^{(n)}(t_j) = \left| b^{(n)}(t_d) \right| e^{in\left(\omega_L t_j - \phi_{0j} - \frac{\pi}{2}\right)}, \qquad 19$$

For better understanding of the modulation phase $\phi_0^j$ that was defined in Eq. 3 for a Gaussian envelope modulated wavepacket, we inspect the expression for the density modulation of an electron passing the PINEM modulation point $z = 0$ at time $t_j$ (See SM Eq. S22):

$$\left| \psi_j(t,z) \right|^2 = v_0 f_{et}\left(t - t_{0j} - \frac{z}{v_0}\right) \left\{ 1 + 2 \sum_n \left| b^{(n)}(z) \right| \cos\left[ n \left[ \omega_L \left( t - t_{0j} - \frac{z}{v_0} \right) + \phi_{0j} + \frac{\pi}{2} \right] \right] \right\} \qquad 20$$

where $f_{et}(t) = \frac{1}{\sqrt{2\pi\sigma_{t0}^2}} \exp\left[-\frac{t^2}{2\sigma_{t0}^2}\right]$ is assumed to be identical for all QEWs, and the assumption is $\sigma_t \gg 2\pi/\omega_L$ or $\sigma_k \ll \delta k$. (the strong recoil regime (26) – the normal regime of PINEM modulation). Also the z-dependent harmonic amplitude $\left| b^{(n)}(t_d) \right|$ of all modulated QEWs is the same as the term in Eq. 14 for all QEWs - $\left| b^{(n)}(z) \right| = J_n\left[ 4|g_L| \sin\left( n \frac{\hbar \delta k^2}{2\gamma^3 m_e} \frac{z}{v_0} \right) \right] e^{-\frac{n^2}{2}\left(\frac{\hbar z}{\gamma^3 m_e v_0}\right)^2 \sigma_k^2 \delta k^2}$, since all experience the same free drift time $z/v_0$ from the PINEM modulation point to z. Both the envelope and the density wave propagate at the same velocity $v_0$, and the phase of the central position of the envelope of the j-th input QEW relative to the laser beam phase $\phi_L$ is determined upon entrance time ($z = 0$, $t = t_{0j}$): $\phi_{0j} = \omega_L t_{0j} - \phi_L$. We may regard this phase as a "Carrier-Envelope phase" - CEP - of the modulated QEW (see SM. 3). Thus,

$$\left| \psi_j(t,z) \right|^2 = v_0 f_{et}\left(t - t_{0j} - \frac{z}{v_0}\right) \left\{ 1 + 2 \sum_n \left| b^{(n)}(z) \right|^2 \cos\left[ n \left[ \omega_L \left( t - \frac{z}{v_0} \right) - \phi_L + \frac{\pi}{2} \right] \right] \right\} \qquad 21$$

Namely, the expectation value of the density modulation function of all QEWs is identical except for the timing of their envelope center (the QEW centroid) (see SM. 3). All modulated QEWs are phase matched at any place, and specifically at the radiative interaction location $z = L_d$.

We define such a beam of multiple modulation-phase matched QEWs as a "modulation-correlated beam". Conversely, if the phases of the bunching factors of the different modulated QEWs are random (e.g. when the modulating light beam is partly coherent and $\phi_L$ is random), the beam is defined as an "uncorrelated modulated QEWs beam".

In Figure 5, we present the multi-electron photon emission growth and the final photon state emitted by a beam of modulated QEWs. Applying the iterative computation procedure of Eq. 12 (illustrated by Fig. 2) for $N_e = 50$ electrons, the final photon state was computed for three cases: a modulation-correlated beam, a non-correlated modulated beam, and an unmodulated beam of QEWs. In the first row, we show the growth of the photon number expectation $\langle n_{ph} \rangle$ with the input electron number. The Wigner distribution of the final quantum emission state is presented in the second row for a single beam emission event. The third row displays the Wigner distribution in the three cases, averaged of multiple independent beam emission events.

The first row of Fig. 5 displays the photon number expectation $\langle n_{ph} \rangle$ in the three cases considered. The orange dots correspond to the computation result according to Eq. 12 and the blue bars represent the results of the analytical formula Eq. 17). Specifically, in the case of a modulation-correlated beam (Fig. a-1), the photon number expectation $\langle n_{ph} \rangle$ grows quadratically with the input electron number $N_e$. This is the hallmark of superradiant emission (33, 34). We compare this result to the analytical expression - Eq. 17). In this case of a modulation-correlated beam, the bunching factors of all the electrons are phase-matched with each other. Setting in (17) $\tilde{b}_i \tilde{b}_j^* = |\tilde{b}_i|^2 = |\tilde{b}^{(1)}|^2$, and counting the number of permutations in the sum: $\binom{N_e}{2}\binom{2}{1} = N_e(N_e - 1)$, the photon number of the superradiant emission is

$$\langle n_{ph} \rangle = N_e |g|^2 + N_e(N_e - 1)|g|^2 |b^{(1)}|^2 \qquad 22$$

This curve is displayed in Fig. 5a by the blue bars, showing full agreement with the computation result. Without modulation ($|b^{(1)}| = 0$), we get the limit of regular spontaneous emission which is proportional to the number of emitting electrons (fig. 5c):

$$\langle n_{ph} \rangle = N_e |g|^2 \qquad 23$$

For the case of a beam of uncorrelated modulated QEWs, the phase matching between the electrons described by Eq. 21 is not satisfied since $\phi_L$ is taken random. Therefore, the photon number grows linearly as a function of the input electrons number $N_e$ with some random deviations as shown by the orange dotted line in Fig. 5b. However, if one performs the same experiment repeatedly for many emission events by multiple QEWs and averages (here we ran the simulation of the interaction with 50 electrons for 1000 times), one finds that $\langle n_{ph} \rangle$ growth with $N_e$ becomes closer to linear, as shown by the blue bars. This linear excitation agrees with Eq. 17 in consideration that $\overline{\sum_{i \neq j}^{N_e} \tilde{b}_i \tilde{b}_j^*} \simeq 0$ because of the randomity of the phases of the bunching factors:

$$\overline{\langle n_{ph} \rangle} = N_e |g|^2 + |g|^2 \overline{\sum_{i \neq j}^{N_e} \tilde{b}_i \tilde{b}_j^*} \simeq N_e |g|^2. \qquad 24$$

The spontaneous emission of the randomly modulated QEWs (Eq. 23) is approximately the same as the spontaneous emission of unmodulated QEWs (Eq. 22) and the slope in Fig. 5b is the same as in Fig. 5c). This is consistent with the argument that spontaneous emission by uncorrelated electrons is independent of the shape or modulation of the individual QEWs (*31*).

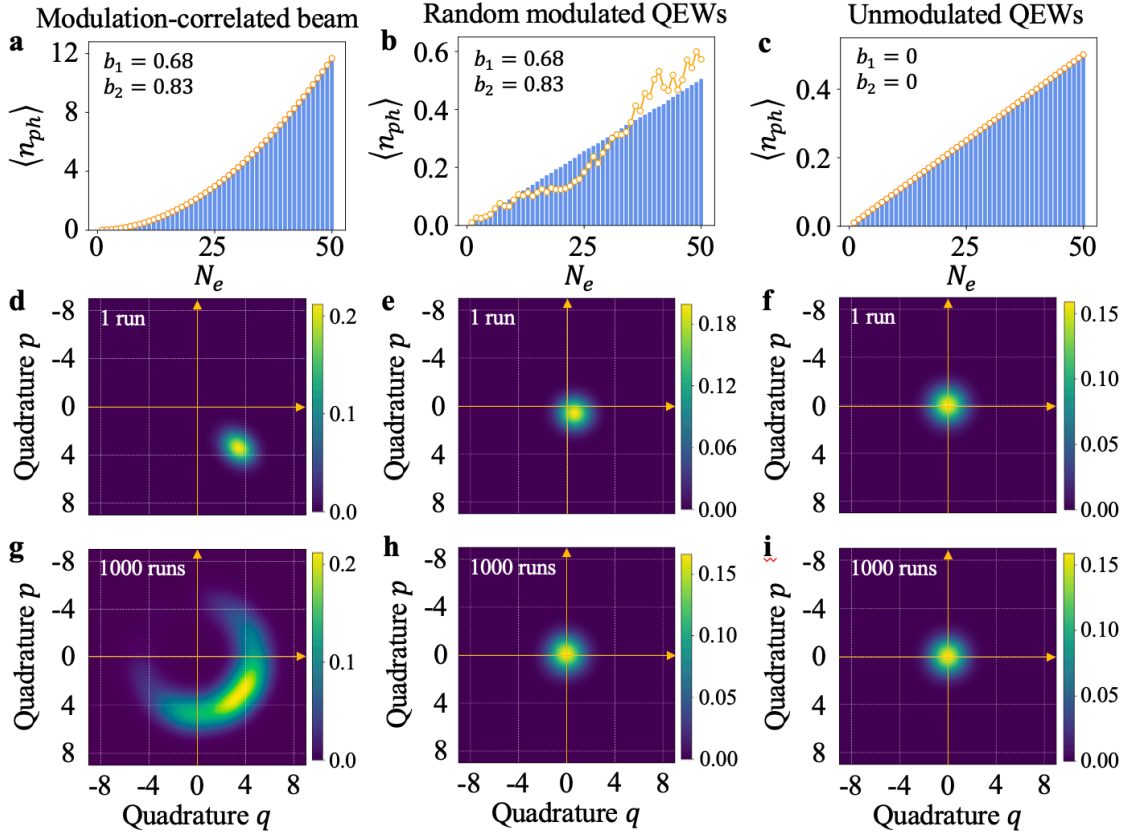

Figure 5: The multi-electron photon emission rate and the final emission state generated by a modulation-correlated beam (column 1), non-correlated modulated beam (column 2) and non-modulated beam (column 3). The dependence of the photon number expectation value $\langle n_{ph} \rangle$ (Eq. 17) on the input electron number $N_e$ is shown in the first row for cases a, b, c. The Wigner distribution of the final emission state event of a beam of $N_e = 50$ electrons is shown in the second row, all calculated for coupling constant $|g| = 0.1$. The third row presents the accumulated Wigner distribution of the photon states generated by multiple beam emission events. In the first row the orange dots correspond to the iterative computation result according to Eq. 12 and the blue bars represent the results of the analytical formula Eq. 17). The bars in subplot b represent averaging over 200 simulation runs (beam emission events).

## Quantum observables in Wigner phase-space distribution

The quadratic growth of superradiant emission by multiple modulation-correlated QEWs (Eq. 22 and Fig. 5a) is agreeable with the classical dependence of superradiant emission in the limit of bunched point particles beam (*34*) (see SM-8). The quantum features of the emission by a single modulated QEW (primarily the off-diagonal elements of the quantum state density matrix) that facilitate the correlated collective (superradiant) emission by multiple modulated QEWs, can only be observed in the Wigner phase-space distribution. The Wigner phase-space distribution is a measurable physical entity. It can be measured by techniques such as homodyne detection (*49*). In rows 2 and 3 of Fig. 5 we show comparatively the expected Wigner distribution patterns of the radiation emission quantum states of a modulation-correlated beam, an uncorrelated modulated beam, and an unmodulated (Gaussian shaped QEWs) beam.

These WD patterns are expected to reveal the nonclassical features of the spontaneous and superradiant emission with the expectation that they can be confirmed experimentally.

The second row in Fig. 5 displays the WD of the final emission quantum state after interaction with a beam of $N_e = 50$ electrons. Fig. 5d corresponds to the case of a modulation-correlated electron beam, namely, all QEWs are PINEM-modulated by the same coherent laser beam of frequency $\omega_L$ and stable phase $\phi_L$ (see Eq. 21). In this case, corresponding to superradiant emission, the center of the WD pattern is displaced off-origin and located at a polar angle corresponding to the phase of the modulating laser. Note also that the pattern shows some photon squeezing features, and the size of the pattern (its variance) contains non-classical information. Fig. 5e displays the WD of the light emitted by the same beam of 50 QEWs that are modulated in this case by an incoherent light source ($\phi_L$ is random). In this case, the WD pattern displays small displacement off-origin at an arbitrary polar angle that relates to the randomicity of the photon growth shown in 5b. Fig. 5f displays the WD of the light emitted by an unmodulated beam (normal spontaneous emission). Its circular symmetric distribution corresponds to super-Poissonion statistics.

The non-classical feature of the radiation emission state of a multiple electron beam can be explained in terms of the analytical expressions for the variance of the WD patterns in the quadrature phase-space presentation. Extending Eq. 15 to multiple electrons, the variance in the x and p dimensions is:

$$\Delta q^2 = \frac{1}{2} + N_e |g|^2 \left(1 - |\tilde{b}^{(1)}|^2\right) - |g|^2 \left(|\tilde{b}^{(1)}|^2 - |\tilde{b}^{(2)}|\right) \sum_{j=1}^{N_e} \cos[2\phi_L]$$
$$\Delta p^2 = \frac{1}{2} + N_e |g|^2 \left(1 - |\tilde{b}^{(1)}|^2\right) + |g|^2 \left(|\tilde{b}^{(1)}|^2 - |\tilde{b}^{(2)}|\right) \sum_{j=1}^{N_e} \cos[2\phi_L]$$
$$\quad 25$$

In the case $|\tilde{b}^{(2)}| \ll |\tilde{b}^{(1)}|^2$ (see SM. 6),

$$\Delta q^2 = \frac{1}{2} + N_e |g|^2 \left(1 - |\tilde{b}^{(1)}|^2\right) - |g|^2 \left(|\tilde{b}^{(1)}|^2\right) \sum_{j=1}^{N_e} \cos[2\phi_L]$$
$$\Delta p^2 = \frac{1}{2} + N_e |g|^2 \left(1 - |\tilde{b}^{(1)}|^2\right) + |g|^2 \left(|\tilde{b}^{(1)}|^2\right) \sum_{j=1}^{N_e} \cos[2\phi_L]$$

The quadrature variables $\Delta q^2$ and $\Delta p^2$ of the WD variance decomposes as a center of mass variance $\Delta_c^2$ that is independent of the modulation phase and is linear with the number of input electron, and a relative variance $\Delta_r^2$ which is a phase dependent term:

$$\Delta_c^2 = \frac{\Delta q^2 + \Delta p^2}{2} = \frac{1}{2} + N_e |g|^2 \left(1 - |b^{(1)}|^2\right)$$
$$\Delta_r^2 = \frac{\Delta q^2 - \Delta p^2}{2} = -|g|^2 |b^{(1)}|^2 \sum_{j=1}^{N_e} \cos[2\phi_L]$$
$$\quad 26$$

Fig. 6 displays the scaling of these two terms as a function of the number of electrons in the beam.

For the case of a correlated beam, the sum in the phase-dependent relative variance $\Delta_r^2$ gives the maximal value $N_e \cos[2\phi_L]$. The phase $\phi_L$ corresponds to the CEP of the first electron in the modulation correlated QEWs beam, while all subsequent QEWs are phase-matched with the first one. This phase determines the polar angle of the WD pattern in Fig. 5d. There it is shown for $\phi_L = -\frac{\pi}{4}$, in which case $\Delta q^2 = \Delta p^2$.

The quadrature WD of the superradiant emission exhibits an apparent shift from the origin:

$$(\langle q \rangle^2 + \langle p \rangle^2)/2 = N_e^2 |g|^2 |b^{(1)}|^2$$

as shown in Fig. 5d, consistently with the prediction of Eq. 18. Note that the WD pattern there displays a squeezing effect, which as predicted by Eq. 25, originates from the $\phi_L$ dependence of the second term in the equation.

As explained next in the discussion section, measuring the WD by the method of homodyne detection requires scanning over the phase of a local oscillator. Therefore, it would be hard to expect the possibility of measuring the entire WD pattern of superradiant emission of a correlated electron beam, as shown in Fig. 5d, in a single radiation emission event. However, assuming a high Q resonator mode, it is conceivable to measure the WD of the accumulated radiation quantum state generated in multiple experiments, as long as the modulating laser beam is coherent, or partly coherent for the duration of the multiple beam radiation events. If the modulating laser phase is stable, the phases of all modulated QEWs will be matched, though their envelope arrival times are random, and all beam radiation events would produce the same WD patterns with the same polar angle $\phi_L$ as shown in Fig. 5d sustained for enough time for scanned homodyne detection.

If the phase of the beam modulation is not stable during the WD measurement process, the polar angle of the WD pattern relative to the origin would contain certain randomness and the accumulated WD will exhibit an arc pattern as shown in Fig. 5g. The extent of the arc depends on the level of laser coherence and consequent phase correlation between the beam radiation events. In the limit of full phase randomicity, the arc pattern would become a ring, but its variance will be still the same. The third row of Fig. 5 exhibits accumulated WD patterns of the photon state emitted by a partly coherent modulated beam for all cases (accumulated with 1000 beam emission events). The arc of the accumulated WD in Fig. 5g is a result of multiple photon emission events (1000) of 50 modulation-correlated QEWs with the initial phase of the laser $\phi_L$ in each event determined statistically as a Gaussian statistical distribution centered at $\pi/4$ with width of $0.3\pi$. The deviation of the arc from origin is proportional to $N_e$, and reflects the superradiant $N_e^2$ growth of the photons number in each 50 electrons emission event in which the modulated QEWs are phase matched to each other.

The Wigner distribution of the final emission of unmodulated QEWs beam is shown in Fig. 5f. Since $b^{(1)} = b^{(2)} = 0$ is valid for all the electrons in the unmodulated beam, the third term in Eq. 17 vanishes and the WD has a circular shape and is centered around the origin of the phase space. Note that in this case $b_i = 0$ and according to Eq. 16 and 18, the density matrix of the final emission state is determined by the coefficient $c_{0,n}^{(0)}$ only (no off-diagonal terms, details in SM. 6):

$$\rho_{ph}^{N_e}(n, n+1) = 0$$

$$\rho_{ph}^{N_e}(n, n) = c_{0,n}^{(0)} = \frac{1}{N_e|g|^2 + 1} \left( \frac{N_e|g|^2}{N_e|g|^2 + 1} \right)^n$$

Namely, the WD of conventional spontaneous emission of unmodulated uncorrelated electrons corresponds to a thermal light source (*51*). The measurable WD shown in Fig. 5i is the same as the WD of single emission event, since $b_i = 0$ for all the electrons involved in the interaction.

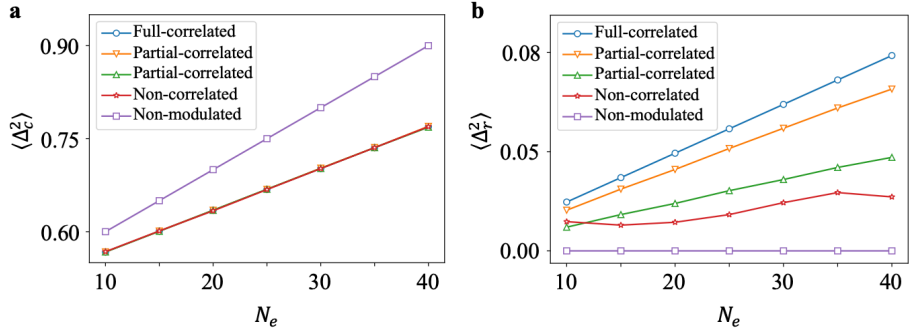

Figure 6: the change of $\Delta_c^2$ and oscillation amplitude of $\Delta_r^2$ with the input electron number $N_e$: (a) the center variance $\langle \Delta_c^2 \rangle$ is always linear to the electron number $N_e$, which is irrelevant to correlation between the input QEWs. (b) the slopes of lines in the figure show the degree of correlation between QEWs. The fully phase-matched QEW beam is shown in blue, while the orange line and green line have the phase distribution in Gaussian, with $\sigma_\phi = 0.3$ and $\sigma_\phi = 0.6$. The variance of the electric field of the non-correlated QEW beam exhibits random fluctuation (red line).

## Discussion

The undertaking of this work is to tackle the debatable issue of spontaneous emission dependence on the shape or modulation of a radiating QEW in the framework of quantum electrodynamics theory (*16, 42, 52*). The QED formulation further enables to understand the possibility of coherent correlated interaction and radiation emission by multiple modulated QEWs, leading, in the case of modulation-correlated QEWs, to the phenomenon of bunched beam superradiance (*24, 34*). In another context, it can lead to enhanced excitation of quantum transitions in a bound electron system (the FEBERI phenomenon (*38*)).

Our QED explanation of the cooperative photon emission process by correlated quantum electron wavepackets, relies on the off-diagonal terms of the radiation quantum density matrix (the quantum state). In the case of a single shaped or modulated QEWs, these off-diagonal terms are not directly measurable and therefore have no physical significance. This explains why the emission probability and photon statistics of spontaneous emission are independent of the shape or modulation of the radiating QEW, as is demonstrated in the first and second columns of Fig. 3 and asserted in earlier theoretical and experimental works (*26, 31, 32*). However, these off-diagonal terms play a cardinal role in establishing a measurable superradiant emission state in the case of modulation-correlated QEWs. The iterative photon state buildup of the radiation quantum state (Eq. 12) that we formulated for the emission by multiple correlated QEWs provides the expected $N^2$ scaling of superradiant emission, as known from classical theory (*33, 34*). This result helps to substantiate the QED formulation that we provide.

For further substantiation of our QED model, we calculated iteratively the radiation quantum state generated by multiple modulation correlated QEWs and presented it in terms of the quadrature Wigner distribution. The Wigner distribution provides full description of the entire quantum density matrix, including off-diagonal terms (*53*). We showed that the WD of the photon state exhibits an imprint of the density modulation of the QEW already at the single QEW case (Figs. 3, 4). The imprint of the modulation on the WD of the photon state generated by multiple correlated QEWs (2nd, 3rd rows in Fig. 5) is even more distinct, and may be measurable. These WD patterns reveal the quantum features that were pointed out in the previous section, including evaluation of the off-diagonal terms of the quantum state.

Measurement of the quadrature WD of a radiation state is possible by the technique of homodyne detection, but this requires scanning over the phase of a local oscillator (*49*). In the scenario of measuring the WD of the radiation emission by multiple modulation-correlated QEWs, as shown in Fig. 5d, this necessitates repeated measurements of emission events by identical ensembles of phase correlated QEWs that were modulated by the same coherent laser beam of stable phase $\phi_L$. This may be attained by phase-locking the local oscillator of the homodyne detection laser to the PINEM modulation laser beam (*48, 49*). If the modulating laser has limited coherence (the phase has finite variance), the measurement of the WD over multiple radiation emission events would result in an arc or a ring distribution as in Fig. 5g.

The thrust of our theory is that the phase of the modulated QEWs in a modulation-correlated electron beam are phase-locked to each other, even *though their arrival time* (the arrival time of their envelope center) to the interaction point *is random*. Yet, the CEP of the first modulated QEW in each multi-electron radiation emission event is still random and depends on its emission time from the cathode. If the electrons are photoemitted from the cathode and their emission times could be controlled better than an optical period by phase locking the photoemission laser beam to the PINEM modulation laser, then repeated measurement of identical WD patterns with the same phase would be possible with multiple radiation events, replicating the exact WD of superradiance shown in Fig. 5d. Furthermore, in this case it would be even possible to construct the WD of the photon state emitted by a single modulated QEW (Fig. 4), because it would correspond to exact replication of the photon state, including phase, in multiple single electron radiation emission events.

Is an optical period timing of photo-electron emission possible in the lab? A recent article of Li et al (*47*) seems to indicate a positive answer, and it opens way to new physical understanding of the photoemission of shaped QEWs from solid. In that work, control was demonstrated over the photoemission current level (or probability of electron emission) relative to the pumping laser beam phase, using a scheme of two-color photoemission. The electron emission time was related there to the phase of the Bloch wavefunction in the solid-state medium of the emitter Tungsten tip. This, in our interpretation, testifies to the role of the finiteness of the Bloch wavepacket in the solid, that retains distinct timing relation to the pumping laser phase even after emission to vacuum. We suggest that the Bloch wavepacket phase in (*47*) is the analogue of the CEP in our model, where the wavepacket envelope is modeled as a Gaussian function. It is questionable whether the photoemission laser can be phase matched to the PINEM modulation laser and maintained during the electron acceleration and transport from the cathode to the modulation point. If this is technically possible, then it is conceivable that superradiant WD of a modulation-correlated QEWs beam (as in Fig. 5d), or even the WD of the radiation by a single modulated QEW (Fig. 4), can be reconstructed by homodyne measurements with repeated phase-locked correlated beam emission events. This is yet an unproven conjectural proposition, if realizable, such measurements may be of interest for substantiation of the QED theory of bunched beam superradiance and its application to other coherent collective interactions of electrons.

## Conclusion

The bridge between classical point-particle description of electron interaction with light and quantum theory plane-wave wavefunction description of such interactions, has been well established for the case of stimulated interaction by invoking the concept of finite size QEW (*26*). Assigning similar dependence to the shape and modulation features of the QEW in the case of spontaneous emission using a semi-classical model (*32*), has limited validity. In fact, QED analysis of spontaneous emission by a single QEW negates the possibility of such dependence (*31*). Yet, in the case of the well-established phenomenon of classical superradiance by a bunched electron beam (*34*), electrons cooperate to contribute collectively to the buildup of the photon emission into a radiation mode. In this work we established, in the framework of QED theory, the mechanism of the cooperative buildup of superradiance by multiple modulation-correlated QEWs. This mechanism relies on the phase of the off-diagonal terms of the density matrix of the quantum radiation state. These terms are immaterial in the case of spontaneous emission from a single electron, but play the crucial role of establishing coherent phase matched superradiant emission by a modulation-correlated QEW beam.

Beyond establishing the consistence of our QED theory with the classical theory of superradiance, predicting quadratic scaling of the photon emission rate with the number of electrons, we provided deeper insight into the QED scheme of QEWs cooperation by examining the Wigner distribution of the emitted radiation state. The measurement of such distribution is experimentally challenging but feasible. We suggest that it would provide further verification of the QED model, reveal the non-classical features of bunched beam superradiance, and open new avenues for utilization of multiple modulation-correlated QEW beams in enhancement of electron interactions with quantum systems, such as FEBERI (*35*, *40*).

In examining the experimental aspects of measuring the WD of superradiant radiation, we shed light on the counterintuitive fact that superradiance can be built-up by modulation-correlated QEWs (modulated by the same coherent laser beam), even though the electrons are emitted *at random* (more precisely – the CEP of the modulated QEWs is random) (*35*, *38*). We showed that quantum state of the collectively emitted radiation carries the modulation characteristics of the modulation-correlated QEW beam and builds up coherently despite the randomicity of the QEWs arrival. We conject that the quantum state of this superradiant emission is measurable by tomographic recording of the Wigner distribution of the state by the technique of homodyne detection, provided that the homodyne local oscillator is phase locked to the modulating laser, and that the laser has coherence time longer than the cavity lifetime. Furthermore, measurement of the state in repeated multiphoton emission events (in a way replicating the state) would be possible, if sub-optical-cycle control of electron photoemission would be realizable (*47*).

Exposing the scheme of controlling the off-diagonal terms of the radiation states of the QEWs through shaping of their bunching (harmonics phase and amplitude control), uncovers a potential to fine-tune quantum light emission properties through the modulation of the QEWs, and opens way for developing exotic new sources of quantum light. This is further supported by earlier demonstrations of coherence (*18*) and of quantum statistics properties (*54*) from the PINEM modulation light to the electron energy spectrum and their radiation emission state. We therefore suggest that the scheme presented here for the phase correlation of quantum light emission, can be extended to production of a variety of quantum radiation sources through engineering the shape and quantum features of QEWs for specific emission. This may have profound implications for applications in quantum communication and sensing.

## Acknowledgments

We acknowledge support by the Israel Science Foundation through grant 1705/22.

Supplementary Materials for

# Shape-Dependence of Spontaneous Photon Emission by Quantum Electron Wavepackets and the QED Origin of Bunched Electron Beam Superradiance

Bin Zhang[1], Reuven Ianconescu[1,2], Aharon Friedman[3], Jacob Scheuer[1], Mikhail Tokman[3], Yiming Pan[4*], Avraham Gover[1,†]

[†]Corresponding author: gover@eng.tau.ac.il

[1] School of Electrical Engineering - Physical Electronics, Center of Laser-Matter Interaction, Tel Aviv University, Ramat Aviv 69978, Israel

[2] Shenkar College of Engineering and Design 12, Anna Frank St., Ramat Gan, Israel

[3] Schlesinger Family Accelerator Centre, Ariel University, Ariel 40700, Israel

[4] ShanghaiTech University, Shanghai, China

**This PDF file includes:**

    Supplementary Text
    Figs. S1 to S5
    Tables S1
    References (1 to 4)

## 1. Derivation of the scattering matrix Eq. (7)

In order to solve the master equation Eq. 1 in the main text, we use the method of the scattering matrix in the interaction picture, which is presented in this section. Since the momentum recoil due to the electron-photon interaction is much smaller than the initial average momentum $p_0$ of the free electron, it is easier to solve the problem in terms of the shifted wave vector $\hbar k = p - p_0$. Then the free electron Hamiltonian could be rewritten as

$$\widehat{H}_e = E_0 + \hbar v_0 \hat{k} + \frac{\hbar^2}{2\gamma^3 m}\hat{k}^2, \tag{S1}$$

The electron state has the form $|\psi_e\rangle = \sum_k c_k |k\rangle$. In terms of the shifted wavenumber k, the coefficients of the PINEM-modulated QEWs (Eq. 3 in the main text) can be written in a short way:

$$c_k = \sqrt{\frac{1}{\sqrt{2\pi}\sigma_k}} \sum_{n=-\infty}^{+\infty} J_n(2g_L) \exp\left[-\frac{(k - n\delta k_L)^2}{4\sigma_k^2} - in\phi_0\right] e^{-iE_k t_d/\hbar}, \tag{S2}$$

where $\delta k_L = \frac{\omega_L}{v_0}$ is the quantum recoil induced by the modulation laser. Accompany with the free Hamiltonian of the cavity mode $\widehat{H}_{ph} = \hbar\omega_c \hat{a}^\dagger \hat{a}$, the free Hamiltonian of the system is

$$\widehat{H}_0 = \widehat{H}_e \otimes \hat{I}_{ph} + \hat{I}_e \otimes \widehat{H}_{ph}. \tag{S3}$$

The interaction Hamiltonian then writes as

$$\widehat{H}_I = -\frac{e}{2\gamma m_e}\left(\hat{p} \cdot \hat{A}(z) + \hat{A}(z) \cdot \hat{p}\right)$$
$$= -\frac{e\hbar}{2\gamma m_e}\left[(\hat{k} + k_0) \cdot \hat{A}(z) + \hat{A}(z) \cdot (\hat{k} + k_0)\right]$$

where the vector potential is $\hat{A}(z) = A_{0,eff} f(z)\left(\hat{a}e^{iq_z z} + \hat{a}^\dagger e^{-iq_z z}\right)$ and the field envelop is a box-like distribution function $f(z) = \left[\tanh\left(\frac{z+L}{0.1}\right) + \tanh\left(-\frac{z-L}{0.1}\right)\right]/2$ shown in Fig. (S1). Thus, the interaction Hamiltonian could be expressed in the tensor product space $|k\rangle \otimes |n\rangle$ as

$$\widehat{H}_I = -\frac{e\hbar A_{0,eff}}{2\gamma m_e}\left[h_I(z) \otimes \hat{a} + h_I^*(z) \otimes \hat{a}^\dagger\right], \tag{S4}$$

where $h_I(z) = (\hat{k} + k_0) \cdot f(z)e^{iq_z z} + f(z)e^{iq_z z} \cdot (\hat{k} + k_0)$. In the interaction picture, the scattering matrix is defined as

$$\hat{S}(t_f, t_i) = \mathcal{T} \exp\left[-\frac{i}{\hbar}\int_{-t_i}^{+t_f} \widehat{H}_I^{int}(t)dt\right], \tag{S5}$$

where time-dependent interaction Hamiltonian is $\hat{H}_I^{int}(t) = e^{i(\hat{H}_e+\hat{H}_{ph})t/\hbar} \hat{H}_I e^{-i(\hat{H}_e+\hat{H}_{ph})t/\hbar}$. Hence, the final density matrix is

$$\rho_{ep}^{f,in}(t_f) = \hat{S}(t_f, t_i)\, \rho_{ep}^{i,in}(t_i)\, \hat{S}^\dagger(t_f, t_i), \qquad (S6)$$

where $\rho_{ep}^{i,in} = \rho_e^{in} \otimes \rho_{ph}^{in}$. Here $\rho_e^{in} = |\psi_e^{in}\rangle\langle\psi_e^{in}|$ is the electron state in the interaction picture, and $|\psi_e^{in}\rangle = e^{i\hat{H}_e t/\hbar}|\psi_e\rangle$. The electron state in the Schrödinger picture is $|\psi_e\rangle = \int dk\, c_k |k\rangle$ (main text Eq. 2). According to Baker–Campbell–Hausdorff formula, $e^{i(\hat{H}_e+\hat{H}_{ph})t/\hbar} = e^{i\hat{H}_e t/\hbar} e^{i\hat{H}_{ph} t/\hbar}$ since $[\hat{k}, \hat{a}^\dagger\hat{a}] = 0$. For a short interaction process, we ignore the small quadratic term in $\hat{k}$ in the free electron Hamiltonian. Then

$$\hat{H}_I^{int}(t) = -\frac{e\hbar A_{0,eff}}{2\gamma m_e}\left[\left(e^{i\hat{H}_e t/\hbar} h_I(z) e^{-i\hat{H}_e t/\hbar}\right) \otimes \left(e^{i\hat{H}_{ph} t/\hbar} \hat{a} e^{-i\hat{H}_{ph} t/\hbar}\right) + h.c.\right]. \qquad (S7)$$

We invoke the following properties of these operators:

$$e^{i\omega_c \hat{a}^\dagger \hat{a} t} \hat{a} e^{-i\omega_c \hat{a}^\dagger \hat{a} t} = \hat{a} e^{-i\omega_c t}$$

$$e^{i\hat{k}z_0} g(\hat{z}) e^{-i\hat{k}z_0} = g(\hat{z} + z_0)$$

where $g(z)$ is an arbitrary z-dependent function, and get

$$\hat{H}_I^{int}(t) = -\frac{\hbar e A_{0,eff}}{2\gamma m_e} \hat{a} e^{-i\omega_c t}\left[(\hat{k}+k_0)\cdot f(z+v_0 t)e^{iq_z(z+v_0 t)} + f(z+v_0 t)e^{iq_z(z+v_0 t)}\cdot(\hat{k}+k_0)\right]$$
$$-\frac{\hbar e A_{0,eff}}{2\gamma m_e} \hat{a}^\dagger e^{i\omega_c t}\left[(\hat{k}+k_0)\cdot f(z+v_0 t)e^{-iq_z(z+v_0 t)} + f(z+v_0 t)e^{-iq_z(z+v_0 t)}\cdot(\hat{k}+k_0)\right]$$

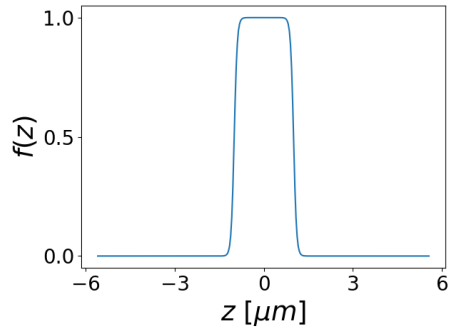

Figure S1: the field distribution function.

Usually, the upper and lower limit $(t_f, t_i)$ of the integration is much smaller than the interaction time, thus we assume $t_f, t_i \to \infty$. Thus, the time integral of $\hat{H}_I^{int}(t)$ in Eq. (S5) produces:

$$\int_{-\infty}^{+\infty} e^{-i\omega_c t} f(z+v_0 t) e^{iq_z(z+v_0 t)} dt = e^{i\frac{\omega_c}{v_0}z} \int_{-\infty}^{+\infty} f(z+v_0 t) e^{i\left(q_z - \frac{\omega_c}{v_0}\right)(z+v_0 t)} dt$$

$$= \frac{2\Lambda}{v_0} e^{i\frac{\omega_c}{v_0}z} \text{sinc}\left[\left(q_z - \frac{\omega_c}{v_0}\right)L\right]$$

which suggests that a synchronism condition $q_z - \frac{\omega_c}{v_0} = 0$ should be satisfied between the free electron and the excited photon mode. Then we get

$$\int_{-\infty}^{+\infty} \widehat{H}_I^{int}(t) dt = -\frac{\hbar e A_{0,eff} L}{\gamma m_e v_0} \hat{a} \left[(\hat{k}+k_0) e^{i\frac{\omega_c}{v_0}z} + e^{i\frac{\omega_c}{v_0}z}(\hat{k}+k_0)\right]$$
$$- \frac{\hbar e A_{0,eff} L}{\gamma m_e v_0} \hat{a}^\dagger \left[(\hat{k}+k_0) e^{-i\frac{\omega_c}{v_0}z} + e^{-i\frac{\omega_c}{v_0}z}(\hat{k}+k_0)\right]$$

According to the commutation relationship $[\hat{z}, \hat{k}] = i$, then

$$\int_{-\infty}^{+\infty} \widehat{H}_I^{int}(t) dt = -\frac{\hbar e A_{0,eff} L}{\gamma m_e v_0} \hat{a} \left[\frac{\omega_L}{v_0} e^{i\frac{\omega_c}{v_0}z} + 2 e^{i\frac{\omega_c}{v_0}z}(\hat{k}+k_0)\right]$$
$$- \frac{\hbar e A_{0,eff} L}{\gamma m_e v_0} \hat{a}^\dagger \left[-\frac{\omega_L}{v_0} e^{-i\frac{\omega_c}{v_0}z} + 2 e^{-i\frac{\omega_c}{v_0}z}(\hat{k}+k_0)\right]$$

$$= -\frac{\hbar e A_{0,eff} L}{\gamma m_e v_0} \left[\left(k_0 + \frac{\omega_L}{2v_0}\right) \hat{a} e^{i\frac{\omega_c}{v_0}z} + \left(k_0 - \frac{\omega_L}{2v_0}\right) \hat{a}^\dagger e^{-i\frac{\omega_c}{v_0}z}\right] - \frac{\hbar e A_{0,eff} L}{\gamma m_e v_0} \left(\hat{a} e^{i\frac{\omega_c}{v_0}z} + \hat{a}^\dagger e^{-i\frac{\omega_c}{v_0}z}\right) \hat{k}$$

Since for a PINEM-modulated electron, the eigenvalue of the wavevector operator $n\frac{\omega_L}{v_0}$ is much smaller than $k_0$: $n\frac{\omega_L}{v_0} \ll k_0$, we have

$$\hat{S}(t_f, t_i) = \hat{S} \simeq \exp(g\hat{a}^\dagger \hat{b} - g^* \hat{a}\hat{b}^\dagger), \tag{S8}$$

where the coupling constant is

$$g = i\frac{eA_{0,eff} k_0 L}{\gamma m_e v_0}, \tag{S9}$$

and the recoil operator is $\hat{b} = e^{-i\frac{\omega_c}{v_0}\hat{z}}$.

## 2. Derivation of the bunching factor of a modulated Gaussian QEW

The n-th order bunching factor is defined in the main text (Eq. 9) as

$$\tilde{b}^{(n)} = \langle \psi_e^{in} | \hat{b}^n | \psi_e^{in} \rangle = e^{i n v_0 \delta k t_0} b^{(n)}, \tag{S10}$$

where $\hat{b}^n = e^{-in\frac{\omega_c}{v_0}\hat{z}} = e^{-in\delta k_c \hat{z}}$ and $|\psi_e^{in}\rangle = e^{iE_k t/\hbar}$ is the free electron state in the interaction picture, $b^{(n)}$ is the bunching factor in the Schrödinger picture.

We can present it in either the coordinate space $\{|z\rangle\}$ or wave vector basis $\{|k\rangle\}$ (k is the shifted wavenumber defined as $k = (p - p_0)/\hbar$)), where $|\psi_e\rangle = \int dz\, \psi_z |z\rangle$ or $\int dk\, c_k |k\rangle$:

$$b^{(n)}(\delta k_c) = \int dk\, dk'\, c_k^* c_{k'} \langle k | \hat{b}^n | k' \rangle = \int dk\, dk'\, c_k^* c_{k'} \langle k | k' - n\delta k_c \rangle = \int dk\, c_k^* c_{k+n\delta k_c}$$

$$b^{(n)}(\delta k_c) = \int dz\, dz'\, \psi_z^* \psi_{z'} e^{-in\delta k_c z} \langle z | z' \rangle = \int dz\, |\psi_z|^2 e^{-in\delta k_c z} \tag{S11}$$

We can also define a spectral distribution function of the QEW density with continuous variable $k$. This is the spatial Fourier transform of the QEW density distribution at any time $t_d$:

$$M_b(k) = \int dz\, |\psi(z, t_d)|^2 e^{-ikz} = \int dk'\, c_{k'}^* c_{k'+k} \tag{S12}$$

Substituting in (S12) the momentum wavefunction of a PINEM modulated QEW $c_k = \frac{1}{\sqrt[4]{2\pi\sigma_k^2}} \sum_{n=-\infty}^{+\infty} J_n(2g_L) \exp\left[-\frac{(k-n\delta k_L)^2}{4\sigma_k^2}\right] e^{-in\phi_0 - iE_k t_d/\hbar}$ (Eq. (S2)), we obtain

$$M_b(k) = \frac{1}{\sqrt{2\pi\sigma_k^2}} \sum_{m',m=-\infty}^{+\infty} J_m^* J_{m'} \int dk'\, \exp\left[-\frac{(k' - m\delta k_L)^2}{4\sigma_k^2} - \frac{(k' - m'\delta k_L + k)^2}{4\sigma_k^2}\right] \\ \times e^{i(m-m')\phi_0} \exp[i(E_{k'} - E_{k'+k})t_d/\hbar] \tag{S13}$$

where $E_{k'} = E_0 + v_0 \hbar k' + \frac{\hbar^2}{2\gamma^3 m_e} k'^2$, is the second order Taylor expansion of the electron's energy. Then the last term gives

$$\exp\left[i\frac{(E_{k'} - E_{k'+k})t_d}{\hbar}\right] = e^{-ikv_0 t_d} \exp\left[i\frac{\hbar t_d}{2\gamma^3 m_e}(k'^2 - (k'+k)^2)\right]$$

$$= e^{-ikv_0 t_d} e^{-i\frac{\hbar t_d}{2\gamma^3 m_e}k^2} \exp\left[-i\frac{\hbar t_d k}{\gamma^3 m_e}k'\right]$$

Hence, the spectral distribution function of the QEW density, $M_b(k)$ becomes:

$$M_b(k) = \frac{1}{\sqrt{2\pi\sigma_k^2}} e^{-ikv_0 t_d} e^{-i\frac{\hbar t_d}{2\gamma^3 m_e}k^2} \sum_{m',m=-\infty}^{+\infty} J_m^* J_{m'} e^{i(m-m')\phi_0} \times$$
$$\int dk' \exp\left[-\frac{(k'-m\delta k_L)^2}{4\sigma_k^2} - \frac{(k'-m'\delta k_L + k)^2}{4\sigma_k^2}\right] e^{-i\frac{\hbar t_d k}{\gamma^3 m_e}k'} \tag{S14}$$

The integration over $k'$ can be treated as the inverse Fourier transform to $z = \frac{\hbar t_d k}{\gamma^3 m_e}$. Performing the Fourier transform:

$$\int dk' \exp\left[-\frac{(k'-m\delta k_L)^2}{4\sigma_k^2} - \frac{(k'-m'\delta k_L + k)^2}{4\sigma_k^2}\right] e^{-ik'z}$$
$$= \sqrt{2\pi\sigma_k^2} \exp\left[-\frac{1}{2}\sigma_k^2 z^2 - i\frac{((m+m')\delta k_L - k)}{2}z\right]$$
$$\times \exp\left[-\frac{1}{8\sigma_k^2}\big((m-m')\delta k_L + k\big)^2\right]$$

and substituting $z = \frac{\hbar t_d k}{\gamma^3 m_e}$, the momentum bunching distribution function is

$$M_b(k) = e^{-ikv_0 t_d} e^{-i\frac{\hbar t_d}{2\gamma^3 m_e}k^2} e^{-\frac{\sigma_k^2}{2}\left(\frac{\hbar t_d}{\gamma^3 m_e}\right)^2 k^2} \sum_{m',m=-\infty}^{+\infty} J_m^* J_{m'} e^{i(m-m')\phi_0} \times$$
$$\exp\left[-i\big((m+m')\delta k_L - k\big)\frac{\hbar t_d k}{2\gamma^3 m_e} - \frac{1}{8\sigma_k^2}\big((m-m')\delta k_L + k\big)^2\right] \tag{S15}$$

By definitions (S9), S10) $b^{(n)}(\delta k_c) \equiv M_b(k = n\delta k_c)$, then we can evaluate the bunching coefficient $b^{(n)}$ (S11) at the cavity frequency $\omega_c$ for a QEW that is PINEM-modulated at frequency $\omega_L$ by substituting $k = n\delta k_c = n\omega_c/v_0$ in $M_b(k)$ (Eq. (S15)). Defining a recoil detuning parameter $\Delta_k = \delta k_L - \delta k_c$ ($\delta k_L = \omega_L/v_0$):

$$b^{(n)}(\delta k_c) = M_b(k = n\delta k_c) = e^{-in\delta k_c v_0 t_d} e^{-in^2\frac{\hbar t_d}{2\gamma^3 m_e}\delta k_c^2} e^{-\frac{n^2}{2}\left(\frac{\hbar t_d}{\gamma^3 m_e}\right)^2 \sigma_k^2 \delta k_c^2} \sum_{m',m=-\infty}^{+\infty} J_m^* J_{m'} e^{i(m-m')\phi_0}$$
$$\times \exp\left[-i\left((m+m'-n) + \frac{n\Delta_k}{\delta k_L}\right) n \frac{\hbar t_d}{2\gamma^3 m_e} \delta k_c \delta k_L\right] e^{-\left((m-m'+n)-\frac{n\Delta_k}{\delta k_L}\right)^2 \frac{\delta k_L^2}{8\sigma_k^2}}$$

The wavepacket duration of a PINEM modulated QEW is much longer than the modulation period $T_L = 2\pi/\omega_L$, and therefore, its momentum spread is much smaller than the laser-induced quantum recoil, namely $\delta k_L \gg \sigma_k$ (large recoil condition) (26). Thus, the last exponential factor diminishes for $m - m' + n \neq 0$ since $\delta k_L^2/8\sigma_k^2 \gg 1$. Keeping in the summation only the terms $m' = m + n$, we obtain

$$b^{(n)}(\delta k_c) = e^{-in(\delta k_c v_0 t_d + \phi_0)} e^{-\frac{n^2}{2}\left(\frac{\hbar t_d}{\gamma^3 m_e}\right)^2 \sigma_k^2 \delta k_c^2} e^{-in^2 \frac{\hbar t_d}{2\gamma^3 m_e}\delta k_c \delta k_L} e^{-\frac{n^2 \Delta_k^2}{8\sigma_k^2}}$$
$$\times \sum_{m=-\infty}^{+\infty} J_m^*(2g_L) J_{m+n}(2g_L) \exp\left[-im\left(\frac{\hbar t_d}{\gamma^3 m_e} n \delta k_c \delta k_L\right)\right] \quad (S16)$$

We apply Graf's addition theorem (*55*) for Bessel functions:

$$\sum_{m=-\infty}^{+\infty} J_m^*(2|g_L|) J_{m+n}(2|g_L|) \exp(im\phi) = e^{in\left(\frac{\pi}{2}-\frac{\phi}{2}\right)} J_n\left(4|g_L|\sin\left(\frac{\phi}{2}\right)\right)$$

with $\phi = -n\frac{\hbar t_d}{\gamma^3 m_e}\delta k_c \delta k_L$ according to Eq. (S16). Then we obtain an analytical expression for the bunching factor of a single modulated QEW interacting with a single radiation mode:

$$b^{(n)}(\delta k_c) = e^{-in\left(\delta k_c v_0 t_d + \phi_0 + \frac{\pi}{2}\right)} e^{-\frac{n^2}{2}\left(\frac{\hbar t_d}{\gamma^3 m_e}\right)^2 \sigma_k^2 \delta k_c^2} J_n\left[4|g_L|\sin\left(n\frac{\hbar t_d}{2\gamma^3 m_e}\delta k_c \delta k_L\right)\right] e^{-\frac{n^2 \Delta_k^2}{8\sigma_k^2}} \quad (S17)$$

The last term shows a fast decay of the bunching factors when the frequencies of the modulating laser and the cavity are not resonant, implying a low efficiency of interaction. For the resonant interaction $(\Delta_k = 0)$ which we are interested in, we denote $\omega_L = \omega_c$ or $\delta k_L = \delta k_c \equiv \delta k$, and get the simple formula

$$b^{(n)}(t_d) = e^{-in\left(\omega_L t_d + \phi_0 + \frac{\pi}{2}\right)} J_n\left[4|g_L|\sin\left(n\frac{\hbar t_d}{2\gamma^3 m_e}\delta k^2\right)\right] e^{-\frac{n^2}{2}\left(\frac{\hbar t_d}{\gamma^3 m_e}\right)^2 \sigma_k^2 \delta k^2} \quad (S18)$$

The amplitude (absolute value) of this formula is identical with the expression earlier developed by [Fan, Abajo] with an additional exponential decay factor in (S16), which originates from the wavepacket chirping during drift. Our Eq. 16 also includes an expression for the phase of $b^{(n)}$. This would be most important for the multi-electron theory of the present work.

## 3. Time-space distribution of the density modulation of a QEW and of a beam of modulation-correlated QEWs

According to Eq. (S12), we can derive the QEW's density distribution at a given drift time $t_d$ through an inverse Fourier transform of $M_b(k)$:

$$|\psi(z,t_d)|^2 = \frac{1}{2\pi}\int dk\, M_b(k) e^{ikz} \qquad (S19)$$

This can be calculated for a PINEM modulated electron by applying the inverse Fourier transform on the expression in S13.

In a PINEM modulated QEW $\delta k_L \gg \sigma_k$ and therefore the Gaussian envelops of the spectral sidebands $\exp\left[-\frac{1}{8\sigma_k^2}\left((m-m')\delta k_L + k\right)^2\right]$ in Eq. (S15) are narrow functions of k and we may use the approximation $k = (m'-m)\delta k_L = n\delta k_L$ in all the coefficients of the sideband Gaussian functions:

$$M_b(k) = e^{-i\frac{\hbar t_d}{2\gamma^3 m_e}(n\delta k_L)^2} e^{-\frac{\sigma_k^2}{2}\left(\frac{\hbar t_d}{\gamma^3 m_e}\right)^2 (n\delta k_L)^2} \sum_{n,m=-\infty}^{+\infty} J_m^* J_{m+n}\, e^{-imn\frac{\hbar t_d}{\gamma^3 m_e}\delta k_L^2}$$
$$\times e^{-in\phi_0} \exp\left[-\frac{1}{8\sigma_k^2}(k - n\delta k_L)^2 - ikv_0 t_d\right]$$

Comparing this expression to the expressions for $b^{(n)}$ ((S16) and (S18)) for $\Delta_k = \delta k_L - \delta k_c = 0$, suggests a simple approximate presentation of the momentum distribution function as a sum of harmonic sidebands of $\omega_L = \omega_c$ or $\delta k \equiv \delta k_L = \delta k_c$:

$$M_b(k) \simeq \sum_n |b^{(n)}(\delta k)| e^{-in\phi_0} \exp\left[-\frac{1}{8\sigma_k^2}(k - n\delta k)^2 - ikv_0 t_d\right] \qquad (S20)$$

Consequently, the density distribution of the bunched QEW is derived from a Fourier transform of (Eq. (S19)):

$$|\psi(z,t_d)|^2 = \sum_n |b^{(n)}(\delta k)| e^{-in\phi_0} \int dk\, \exp\left[-\frac{1}{8\sigma_k^2}(k - n\delta k)^2\right] e^{ik(z-v_0 t_d)}$$
$$= \frac{1}{\sqrt{2\pi\sigma_{z0}^2}} \exp\left[-\frac{(z-v_0 t_d)^2}{2\sigma_{z0}^2}\right] \sum_n |b^{(n)}(\delta k)| e^{-in\phi_0} e^{in\delta k(z-v_0 t_d)} \qquad (S21)$$

where $\sigma_{z0} = 1/2\sigma_k$ is the spatial wavepacket size at the waist. Here we ignored the chirping effect (that has been considered in (26)) since $\sigma_{z0}$ is large. Thus, the density distribution of a single modulated QEW can be presented as a carrier wave composed of spatio-temporal harmonics of bunching amplitude $b^{(n)}(\delta k)$ propagating at the electron average velocity $v_0$ bounded by a QEW Gaussian envelop (normalized in the spatial dimension):

$$f_e(z - v_0 t_d) = \frac{1}{\sqrt{2\pi\sigma_{z0}^2}} \exp\left[-\frac{(z - v_0 t_d)^2}{2\sigma_{z0}^2}\right]$$

Using Eq. (S18) and the relation $J_n(x) = J_{-n}(-x)$, the density bunching distribution can be presented by the following transparent formula:

$$|\psi(z,t_d)|^2 = f_e(z - v_0 t_d) \sum_{n=-\infty}^{+\infty} b^{(n)}(\delta k)\, e^{in\delta k_L z}$$

$$= f_e(z - v_0 t_d)\left[b^{(0)} + \sum_{n=1}^{+\infty}\left(b^{(n)}(\delta k) e^{in\delta k_L z} + b^{(-n)}(\delta k) e^{-in\delta k_L z}\right)\right] \quad (S22)$$

$$= f_e(z - v_0 t_d)\left\{1 + 2\sum_n |b^{(n)}| \cos\left[n\left[\delta k(z - v_0 t_d) - \phi_0 + \frac{\pi}{2}\right]\right]\right\}$$

where $|b^{(n)}(t_d)| = J_n\left[4|g_L|\sin\left(n\frac{\hbar \delta k^2}{2\gamma^3 m_e} t_d\right)\right] e^{-\frac{n^2}{2}\left(\frac{\hbar t_d}{\gamma^3 m_e}\right)^2 \sigma_k^2 \delta k^2}$ is the amplitude of $b^{(n)}$. Note that at $t_d = 0$, right after the momentum modulation point, there is no density modulation - $|\psi(z,t_d)|^2 = f_e(z - v_0 t_d)$, and the density modulation develops only after a finite drift time up to the interaction region origin $t_d = L_d/v_0$.

The density modulation of the QEW in terms of time harmonics of the modulation frequency $\omega_L = \omega_c$ at a drift distance $z_d = v_0 t_d$ away from the PINEM modulation point:

$$|\psi(t, L_d)|^2 = v_0 f_e\left(t - \frac{z_d}{v_0}\right)\left\{1 + 2\sum_n |b^{(n)}(t_d)| \cos\left[n\left[\omega_L\left(t - \frac{L_d}{v_0}\right) + \phi_0 + \frac{\pi}{2}\right]\right]\right\} \quad (S23)$$

In figure S2 we display the density distribution in terms of space (z) at a fixed time $t_d$. There is excellent agreement between the exact waveform derived by direct calculation of $|\psi(z,t_d)|^2$ where $\psi(z,t_d)$ is the inverse Fourier transforms of $c_k$ (Eq. S2) or by inverse Fourier transform of $M_b(k)$ (Eqs. 17, 13) shown in the top panel of Fig. S2 and the approximate expression Eq. (S22) shown in the bottom panel.

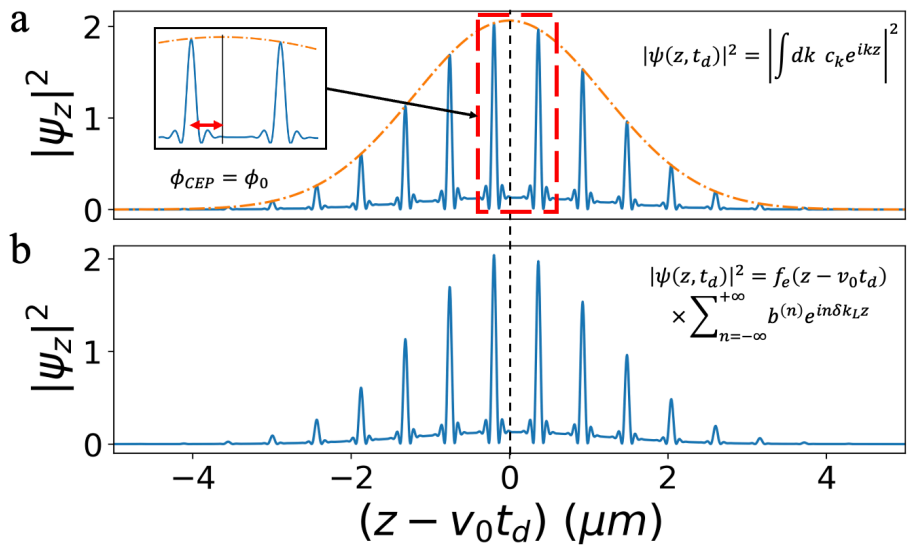

Figure. S2: Comparison of the density distribution of a modulated QEW of: (a) an exact expression based on direct calculation of $|\psi(z,t_d)|^2$ where $\psi(z,t_d)$ is the inverse Fourier transforms of $c_k$ (Eq. S2) and (b) the approximate expression Eq. (S22). This was calculated for $\sigma_t = 1.5T$, $g_L = 1.5$ and $L_d = 3\ cm$. The modulation phase $\phi_0 = -0.2\pi$, which is exactly the CEP of the bunched QEW as shown in the subplot in (a).

For multiple modulated QEWs arriving at the modulation point at random time, we can write the density distribution of j-th input QEW as function of (z, t) according to Eq. (S23)

$$|\psi_j(t,z)|^2 = v_0 f_{et}\left(t - t_{0j} - \frac{z}{v_0}\right)\left\{1 + 2\sum_n \left|b^{(n)}\left(\frac{z}{v_0}\right)\right|\cos\left[n\left[\omega_L\left(t - t_{0j} - \frac{z}{v_0}\right) + \phi_{0j} + \frac{\pi}{2}\right]\right]\right\} \quad (S24)$$

where $t_{0j}$ is the time when the center of the envelope of j-th electron enters the modulation point $z = 0$. Since the interaction happens at $z = L_d$ for all the electrons, the position of carrier wave's peak is determined by $\cos\left[n\left[\omega_L\left(t - t_{0j} - \frac{L_d}{v_0}\right) + \phi_{0j} + \frac{\pi}{2}\right]\right] = 1$, hence $\omega_L\left(t - t_{0j} - \frac{z_d}{v_0}\right) + \phi_{0j} + \frac{\pi}{2} = 2N_j\pi$ and the arrival time of the peak at the interaction region is

$$t = \frac{L_d}{v_0} - \frac{\pi}{2\omega_L} + \frac{1}{\omega_L}(\omega_L t_{0j} - \phi_{0j} + 2N_j\pi)$$

where $N_j$ is an arbitrary integer number. As long as the term $\omega_L t_{0j} - \phi_{0j} = \phi_L$ is constant, the peaks of the carrier wave is a pulse-train in time with the period $2\pi/\omega_L$. Since the electrons move at the same speed $v_0$, the the position of the micro-bunches formed within the density-modulated QEW in the spatial domain is

$$z_{\mu b} = \frac{v_0}{\omega_L}\left(2\pi N_j + \phi_L - \frac{\pi}{2}\right) + L_d \quad (S25)$$

as shown in the dashed box in Fig. S3.

According to Eq. (S18), the bunching factor of the QEWs experience the same drift distance $z_d$ in the Schrodinger picture is

$$b_j^n(t_d) = e^{-in\left(\phi_j + \omega_L t_d + \frac{\pi}{2}\right)} J_n\left[4|g_L|\sin\left(n\frac{\hbar\delta k^2}{2\gamma^3 m_e}t_d\right)\right] e^{-\frac{n^2}{2}\left(\frac{\hbar t_d}{\gamma^3 m_e}\right)^2 \sigma_k^2 \delta k^2}$$

While in the interaction picture, the bunching factor is

$$\tilde{b}_j^{(n)} = e^{in\delta k(t_{0j}+t_d)}b^{(n)}(t_d) = |b^{(n)}(t_d)|\, e^{in\left(\omega_L t_{0j} - \phi_j - \frac{\pi}{2}\right)} \quad (S26)$$

which is consistent with Eq. (S24)

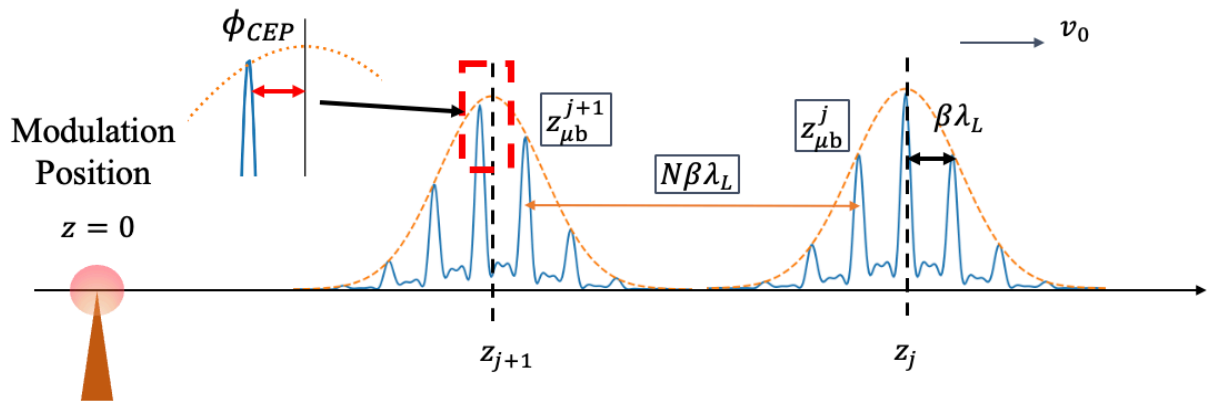

Figure S3: The j-th and (j+1)-th input QEWs after modulation. $z_j$ and $z_{j+1}$ are the position of j-th and (j+1)-th electron (their envelope centers) at any given time past modulation. $z_{\mu b}^{j}$ and $z_{\mu b}^{j+1}$ are the positions of microbunches of the two QEWs when the QEWs are "modulation-correlated" (Eq. (S25)).

## 4. Deviation of equation 8

Due to the recoil operator $\hat{b} = e^{-i\frac{\omega_c}{v_0}\hat{z}}$ satisfies $[\hat{b}^\dagger, \hat{b}] = 0$, the scattering matrix Eq. (S8) can be expanded (56) as

$$\hat{S} = \exp(g\hat{a}^\dagger \hat{b} - g^* \hat{a}\hat{b}^\dagger)$$
$$= \exp(g\hat{a}^\dagger \hat{b}) \exp(-g^* \hat{a}\hat{b}^\dagger) \exp\left(-\frac{1}{2}[g\hat{a}^\dagger \hat{b}, -g^* \hat{a}\hat{b}^\dagger]\right)$$
$$= \exp\left(-\frac{|g|^2}{2}\right) \sum_{n,m} \frac{1}{n!}\frac{1}{m!} g^n (-g^*)^m (\hat{a}^\dagger \hat{b})^n (\hat{a}\hat{b}^\dagger)^m \quad \text{(S27)}$$
$$= \exp\left(-\frac{|g|^2}{2}\right) \sum_{n,m} \frac{1}{n!}\frac{1}{m!} g^n (-g^*)^m (\hat{a}^\dagger)^n (\hat{a})^m \hat{b}^{n-m}$$

where we use the relation $(\hat{b}^\dagger)^m = \hat{b}^{-m}$. Thus, we could compute the general form of transition matrix element between the combined basis $|k_i, n_i\rangle$ and $|k_f, n_f\rangle$ for the electron and photon state (main text Eq. ()):

$$\langle k_f, n_f | \hat{S} | k_i, n_i \rangle = e^{-\frac{|g|^2}{2}} \sum_{n,m} \frac{g^n}{n!} \frac{(-g^*)^m}{m!} \langle k_f | \hat{b}^{n-m} | k_i \rangle \otimes \langle n_f | (\hat{a}^\dagger)^n (\hat{a})^m | n_i \rangle. \quad \text{(S28)}$$

The recoil operator gives

$$\langle k_f | \hat{b}^{n-m} | k_i \rangle = \langle k_f | k_i - (n-m)\delta k \rangle = \delta_{k_f, k_i - (n-m)\delta k}, \quad \text{(S29)}$$

and the photon generation/annihilation operator gives

$$\langle n_f | (\hat{a}^\dagger)^n (\hat{a})^m | n_i \rangle = \sqrt{\frac{n_i!}{(n_i - m)!}} \sqrt{\frac{(n_i - m + n)!}{(n_i - m)!}} \langle n_f | n_i - m + n \rangle$$
$$= \sqrt{\frac{n_i!}{(n_i - m)!}} \sqrt{\frac{(n_i - m + n)!}{(n_i - m)!}} \delta_{n_f, n_i - m + n} \quad \text{(S30)}$$

where $m$ satisfies $m \leq n_i$. Substitude Eq. (S29), (S30) into Eq. (S28), we get

$$\langle k_f, n_f | \hat{S} | k_i, n_i \rangle = \delta_{k_f + n_f \delta k, k_i + n_i \delta k} M_{n_f, n_i}, \quad \text{(S31)}$$

where the delta function $\delta_{k_f + n_f \delta k, k_i + n_i \delta k}$ constrains the momentum conservation of the system, and the transition matrix is

$$M_{n_f, n_i} = e^{-\frac{|g|^2}{2}} g^{n_f - n_i} \sqrt{n_i! n_f!} \sum_{m=\max\{0, n_i - n_f\}}^{n_i} \frac{(-|g|^2)^m}{m!(m + n_f - n_i)!(n_i - m)!}, \quad \text{(S32)}$$

which satisfies $M_{n_i,n_f} = (-1)^{n_f-n_i} M^*_{n_f,n_i}$. According to Eqs. (S30) and (S32), we have found the transition matrix $M$ is purly denpendent on the initial and final photon state, and the coupling constant $g$, which can be rewritten as

$$M_{n_f,n_i} = \langle n_f | \hat{D}(g) | n_i \rangle, \tag{S33}$$

where $\hat{D}(g) = \exp(g\hat{a}^\dagger - g^*\hat{a})$ is the displacement operator for the photon state and the generator of a coherent state $\hat{D}(g)|0\rangle = |g\rangle$. Hence for a general initial state in Schrödinger picture

$$\hat{\rho}^{i,S}_{ep} = |\Psi_i\rangle\langle\Psi_i| = \sum_{n,n'} \rho^i_{ph}(n,n') \sum_{kk'} c_k c^*_{k'} |k,n\rangle\langle k',n'|.$$

The initial combined state of electron and photon in the interaction picture is defined as

$$\begin{aligned}
\hat{\rho}^{i,in}_{ep}(t_i) &= e^{i\hat{H}_0 t_i/\hbar} \hat{\rho}^{i,S}_{ep} e^{-i\hat{H}_0 t_i/\hbar} \\
&= \sum_{n,n'} \rho^i_{ph}(n,n') \sum_{kk'} c_k c^*_{k'} e^{i\hat{H}_0 t_i/\hbar} |k,n\rangle\langle k',n'| e^{-i\hat{H}_0 t_i/\hbar} \\
&= \sum_{n,n'} \rho^i_{ph}(n,n') \sum_{kk'} c_k c^*_{k'} e^{i(E_k+n\hbar\omega)t_i/\hbar} |k,n\rangle\langle k',n'| e^{-i(E_{k'}+n'\hbar\omega)t_i/\hbar} \\
&= \sum_{n,n'} \tilde{\rho}^{i,in}_{ph}(n,n',t_i) \sum_{kk'} e^{iE_k t_i/\hbar} c_k c^*_{k'} e^{-iE_{k'}t_i/\hbar} |k,n\rangle\langle k',n'|
\end{aligned} \tag{S34}$$

where $\tilde{\rho}^i_{ph}(n,n',t_i) = \rho^i_{ph}(n,n') e^{i(n-n')\omega_c t_i}$ is the photon state in interaction picture and $c_k$ is the electron state in Schrodinger picture. Since in the first section, the $k^2$ term is ignorable in the interaction process, we obtain

$$\hat{\rho}^{i,in}_{ep}(t_i) = \sum_{n,n'} \tilde{\rho}^i_{ph}(n,n',t_i) \sum_{kk'} c_k c^*_{k'} e^{iv_0(k-k')t_i} |k,n\rangle\langle k',n'| \tag{S35}$$

The final combined state after interaction is

$$\hat{\rho}^{f,in}_{ep}(t_f) = \hat{S} \hat{\rho}^{i,in}_{ep}(t_i) \hat{S}^\dagger. \tag{S36}$$

The time dependence of the scattering matrix can be ignored when $t_f, t_i \to \infty$ are much larger than the interaction region. Thus, the matrix element of the final photon state can be derived through tracing out the electron's degree of freedom

$$\tilde{\rho}^f_{ph}(n_1,n_2,t_f) = \langle n_1 | \text{Tr}_e\{\hat{\rho}^{f,in}_{ep}\} | n_2 \rangle = \sum_{k_t} \langle k_t, n_1 | \hat{\rho}^{f,in}_{ep} | k_t, n_2 \rangle. \tag{S37}$$

It's worth mention that the photon state derived through partial tracing is in the interaction picture. According to Eq. (S35) and (S36), we can get

$$\tilde{\rho}_{ph}^{f}(n_1, n_2, t_f) = \sum_{k_t} \sum_{n,n'} \tilde{\rho}_{ph}^{i}(n, n', t_i) \sum_{kk'} c_k c_{k'}^* \, e^{iv_0(k-k')t_i} \langle k_t, n_1|\hat{S}|k, n\rangle \langle k', n'|\hat{S}^\dagger|k_t, n_2\rangle. \tag{S38}$$

With Eq. (S31), we know

$$\langle k_t, n_1|\hat{S}|k, n\rangle = \delta_{k_t+n_1\delta k, k+n\delta k} \, M_{n_1, n}$$
$$\langle k_t, n_2|\hat{S}^\dagger|k', n'\rangle = \delta_{k_t+n_2\delta k, k'+n'\delta k} M_{n', n_2}^\dagger$$

then Eq. (S38) can be further simplified

$$\tilde{\rho}_{ph}^{f}(n_1, n_2) = \sum_{n,n'} \tilde{\rho}_{ph}^{i}(n, n') \, M_{n_1, n} M_{n', n_2}^\dagger \sum_{k_t} \sum_{kk'} c_k c_{k'}^*$$

$$\times \delta_{k_t+n_1\delta k, k+n\delta k} \delta_{k_t+n_2\delta k, k'+n'\delta k} e^{iv_0(k-k')t_i}$$

$$= \sum_{n,n'} \tilde{\rho}_{ph}^{i}(n, n') \, M_{n_1, n} M_{n', n_2}^\dagger \sum_{k_t} \sum_{k} c_k c_{k+(n_2-n')\delta k}^* \, \delta_{k_t+n_1\delta k, k+n\delta k} \, e^{iv_0\left(k-(k_t+n_2\delta k-n'\delta k)\right)t_i}$$

$$= \sum_{n,n'} \tilde{\rho}_{ph}^{i}(n, n') \, M_{n_1, n} M_{n', n_2}^\dagger \sum_{k} c_k c_{k+(n_2-n_1+n-n')\delta k}^* \, e^{iv_0(n_2-n_1+n-n')\delta k t_i} \tag{S39}$$

where the latter summation of the wavefunction $c_k$ (Schrödinger picture) over $k$ gives the QEW's bunching factor with order $(n_1 - n) - (n_2 - n')$, thus

$$\tilde{\rho}_{ph}^{f}(n_1, n_2) = \sum_{n,n'} \tilde{\rho}_{ph}^{i}(n, n') \, \tilde{b}^{\left((n_1-n)-(n_2-n')\right)} M_{n_1, n} M_{n', n_2}^\dagger, \tag{S40}$$

where $\tilde{b}^{(n)} = b^{(n)} e^{iv_0 n \delta k t_i}$ is the bunching factor in the interaction picture. For the interaction with j-th input electron, assume the remained photonic state in the cavity is $\tilde{\rho}_{ph}^{j-1}$, then we have

$$\tilde{\rho}_{ph}^{j}(n_j, n_j') = \sum_{r_j=-\infty}^{\infty} \tilde{b}_j^{(r_j)} \sum_{n} \tilde{\rho}_{ph}^{j-1}(n, n+r_j+n_j'-n_j) \, M_{n_j, n} M_{n+r_j+n_j'-n_j, n_j'}^\dagger$$

where $\tilde{b}_j^{(r_j)}$ is the r-th order bunching factor for j-th electron. Define $s_j \equiv n_j' - n_j$, then we get:

$$\tilde{\rho}_{ph}^{j}(n_j, n_j+s_j) = \sum_{r_j=-\infty}^{\infty} \tilde{b}_j^{(r_j)} \sum_{n} \tilde{\rho}_{ph}^{j-1}(n, n+r_j+s_j) \, M_{n_j, n} M_{n+r_j+s_j, n_j+s_j}^\dagger. \tag{S41}$$

## 5. Analysis of the photonic state of spontaneous emission by a single QEW

For the first interaction, the cavity is in a vacuum state and the combined state is $|\Psi_i\rangle = |\psi_e\rangle \otimes |0\rangle$. if we assume the initial time of the interaction $t_i = 0$, yielding the state in interaction picture

$$\tilde{\rho}_{ph}^0(n, n+r+s) = \rho_{ph}^0(n, n+r+s)e^{-i(r+s)\omega_c t_i} = \delta_{n,0}\delta_{n+r+s,0}. \tag{S42}$$

Thus, Eq. (S41) can be simplified as

$$\rho_{ph}^1(n_f, n_f + s) = \sum_{r=-\infty}^{\infty} \tilde{b}^{(r)} \sum_n \delta_{n,0}\delta_{n+r+s,0} M_{n_f,n} M^\dagger_{n+r+s,n_f+s}$$
$$= \tilde{b}^{(-s)} M_{n_f,0} M^\dagger_{0,n_f+s}$$

where $M_{n_f,0} = \langle n_f|g\rangle$ and $M^\dagger_{0,n_f+s} = \langle n_f + s|g\rangle$, and $\tilde{b}^{(-s)}$ is the bunching factor of the first input electron.

This result can be obtained in a different manner. As the initial state is vacuum, the annihilation operators in Eq. (S27) can be neglected, yielding

$$\left|\Psi_f^{(1)}\right\rangle = \hat{S}\left|\Psi_i^{(1)}\right\rangle = \exp\left(-\frac{|g|^2}{2}\right) \sum_n \frac{1}{n!} g^n (\hat{a}^\dagger)^n \hat{b}^n |\psi_e\rangle \otimes |0\rangle$$
$$= \exp\left(-\frac{|g|^2}{2}\right) \sum_n \frac{g^n}{\sqrt{n!}} \hat{b}^n |\psi_e\rangle \otimes |n\rangle$$

The corresponding density matrix is, therefore,

$$\hat{\rho}_{f,ep}^{(1)} = \left|\Psi_f^{(1)}\right\rangle\left\langle\Psi_f^{(1)}\right| = \exp(-|g|^2) \sum_{n,m} \frac{g^n}{\sqrt{n!}} \frac{(g^*)^m}{\sqrt{m!}} \hat{b}^n |\psi_e\rangle\langle\psi_e|\hat{b}^{-m} \otimes |n\rangle\langle m|$$

The final photon state after the first interaction is obtained through partial tracing,

$$\hat{\rho}_{ph}^{(1)} = \text{Tr}_e\left(\hat{\rho}_{f,ep}^{(1)}\right) = \exp(-|g|^2) \sum_{n,m} \frac{g^n}{\sqrt{n!}} \frac{(g^*)^m}{\sqrt{m!}} \text{Tr}_e\left(\hat{b}^n |\psi_e\rangle\langle\psi_e|\hat{b}^{-m}\right)|n\rangle\langle m|$$
$$= \exp(-|g|^2) \sum_{n,m} \frac{g^n}{\sqrt{n!}} \frac{(g^*)^m}{\sqrt{m!}} \langle\psi_e|\hat{b}^{n-m}|\psi_e\rangle |n\rangle\langle m|$$

Introduce the bunching factor (or coherence factor) for different orders $b^{(n)} = \langle\psi_e|\hat{b}^n|\psi_e\rangle$ leads to a simplified expression

$$\hat{\rho}_{ph}^{(1)} = \exp(-g^2) \sum_{n,m} \tilde{b}^{(n-m)} \frac{g^n}{\sqrt{n!}} |n\rangle\langle m| \frac{(g^*)^m}{\sqrt{m!}}. \tag{S43}$$

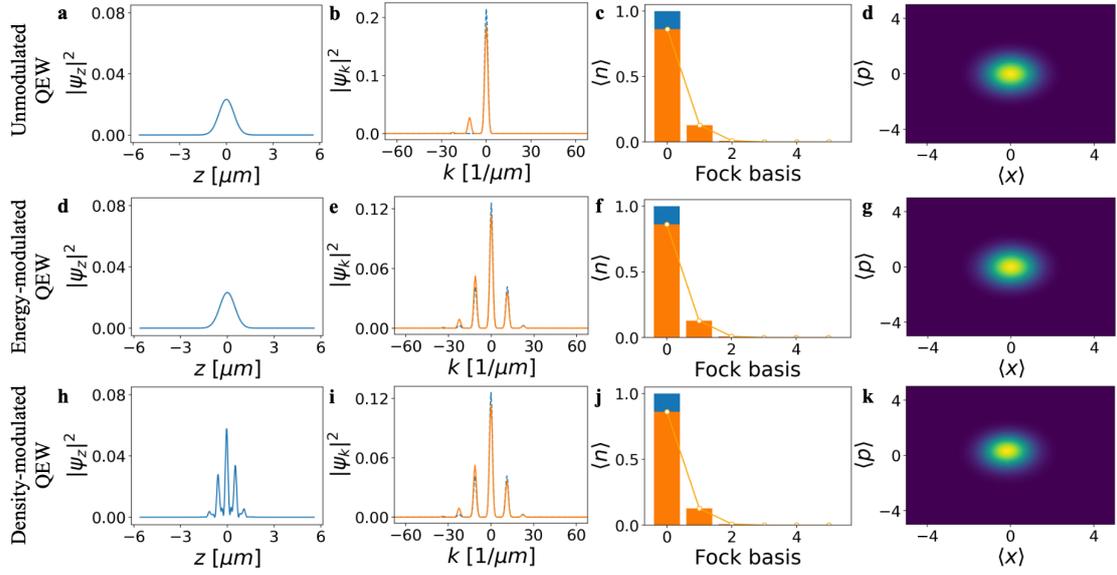

Figure S4: Simulation results of the first interaction for (a) unmodulated, (d) energy-modulated and (h) density-modulated QEW with the vacuum cavity. (b)(e)(i) the momentum distribution of the final (orange line) initial (dotted blue line) electron states. (c)(f)(j) - the final photonic state in Fock space, (d)(g)(k) - the corresponding Wigner distribution. (the results of g=0.39)

## 6. Iterative derivation of the photon state evolvement due to correlated spontaneous emission by multiple QEWs

As shown in Eq. (S41), there are contributions from different orders of the bunching factor for one interaction process. While for one element $\tilde{\rho}_{ph}^{j}(n_j, n_j + s_j)$ of the final photon state, the contribution from the $(r_j + 1)$-th order bunching factor $\tilde{b}_j^{(r_j+1)} M(n + r_j + 1 + s_j, n_j + s_j)$ (assume $r_j \geq 0$) will be approximately $\sim g \tilde{b}_j^{(r_j+1)} M(n + r_j + s_j, n_j + s_j)$. In an experimental situation, the coupling constant is usually very small (on the magnitude of $\sim 10^{-2}$), thus the dominant contribution to the interaction comes from the 1$^{st}$ order bunching factor $b^{(1)}$ after multiple interactions (this is also confirmed by a numerical simulation). Then we have a strong limitation condition for $r_j$: $r_j = -1, 0, 1$, leads to

$$\tilde{\rho}_{ph}^{j}(n_j, n_j + s_j) = \sum_{r_j=-1}^{1} \tilde{b}_j^{(r_j)} \sum_{n} \tilde{\rho}_{ph}^{j-1}(n, n + r_j + s_j) M_{n_j,n} M_{n+r_j+s_j,n_j+s_j}^{\dagger} \tag{S44}$$

The spontaneous emission starts from the vacuum state $|0\rangle$,

$$\tilde{\rho}_{ph}^{0}(n_0, n_0') = \delta_{n_0,0} \delta_{n_0',0}$$

According to Eq. (S44), the photon state after interaction with the first electron is

$$\tilde{\rho}_{ph}^{1}(n_1, n_1 + s_1) = \sum_{r_1=-1}^{1} \tilde{b}_1^{(r_1)} \sum_{n_0} \tilde{\rho}_{ph}^{0}(n_0, n_0 + r_1 + s_1) M_{n_1,n_0} M_{n_0+r_1+s_1,n_1+s_1}^{\dagger}. \tag{S45}$$

Substituting the initial photon state $\tilde{\rho}_{ph}^{0}(n_0, n_0') = \delta_{n_0,0} \delta_{n_0',0}$ into the equation, results in

$$\tilde{\rho}_{ph}^{1}(n_1, n_1 + s_1) = \sum_{r_1=-1}^{1} \tilde{b}_1^{(r_1)} \delta(r_1 + s_1) M_{n_1,0} M_{0,n_1+s_1}^{\dagger}$$

where $M_{n_1,0} M_{0,n_1+s_1}^{\dagger} = \langle n_1|g\rangle\langle g|n_1 + s_1\rangle$ according to Eq. (S33). The photon state after interaction with the second electron is

$$\tilde{\rho}_{ph}^{2}(n_2, n_2 + s_2) = \sum_{r_2=-1}^{1} \tilde{b}_2^{(r_2)} \sum_{n_1} \tilde{\rho}_{ph}^{1}(n_1, n_1 + r_2 + s_2) M_{n_2,n_1} M_{n_1+r_2+s_2,n_2+s_2}^{\dagger}, \tag{S46}$$

replacing $s_1$ in Eq. (S45) with $r_2 + s_2$, and substituting the reformed equation into Eq. (S46), results in

$$\tilde{\rho}_{ph}^2(n_2, n_2 + s_2) = \sum_{r_1, r_2 = -1}^{1} \tilde{b}_1^{(r_1)} \tilde{b}_2^{(r_2)} \delta(r_1 + r_2 + s_2) \sum_{n_1} M_{n_2, n_1} M_{n_1, 0} \\ \times M_{0, n_1 + r_2 + s_2}^\dagger M_{n_1 + r_2 + s_2, n_2 + s_2}^\dagger ,$$ (S47)

Again, for the photon state after interaction with the third electron,

$$\tilde{\rho}_{ph}^3(n_3, n_3 + s_3) = \sum_{r_3 = -1}^{1} \tilde{b}_3^{(r_3)} \sum_{n_2} \tilde{\rho}_{ph}^2(n_2, n_2 + r_3 + s_3) M_{n_3, n_2} M_{n_2 + r_3 + s_3, n_3 + s_3}^\dagger$$ (S48)

after replacing $s_2$ in Eq. (S47) with $r_3 + s_3$ and substituting it into Eq. (S48),

$$\tilde{\rho}_{ph}^3(n_3, n_3 + s_3) = \sum_{r_1, r_2, r_3 = -1}^{1} \tilde{b}_1^{(r_1)} \tilde{b}_2^{(r_2)} \tilde{b}_3^{(r_3)} \delta(r_1 + r_2 + r_3 + s_3) \\ \sum_{n_2, n_1} M_{n_3, n_2} M_{n_2, n_1} M_{n_1, 0} M_{0, n_1 + r_2 + r_3 + s_3}^\dagger M_{n_1 + r_2 + r_3 + s_3, n_2 + r_3 + s_3}^\dagger M_{n_2 + r_3 + s_3, n_3 + s_3}^\dagger$$ (S49)

Thus, we could get the general form of the photon state $\tilde{\rho}_{ph}^{N_e}$ after the interaction with $N_e$ electrons using the mathematical induction method,

$$\tilde{\rho}_{ph}^{N_e}(n_{N_e}, n_{N_e} + s_{N_e}) = \sum_{r_1, r_2, \ldots, r_{N_e}} \left( \prod_{i=1}^{N_e} \tilde{b}_i^{(r_i)} \right) \delta\left( \sum_{j=1}^{N_e} r_j + s_{N_e} \right) \sum_{n_i}^{\{i=0,1,2,\ldots,N_e-1\}} \delta_{n_0, 0} \\ \times \prod_{i=1}^{N_e} M_{n_i, n_{i-1}} \prod_{i=0}^{N_e-2} M_{n_i + \sum_{j=i+1}^{N_e} r_j + s_{N_e}, n_{i+1} + \sum_{j=i+2}^{N_e} r_j + s_{N_e}}^\dagger M_{n_{N_e-1} + r_{N_e} + s_{N_e}, n_{N_e} + s_{N_e}}^\dagger$$ (S50)

As shown in the above equation, the $s_{N_e}$-th diagonal term of the final photon state's density matrix can be expressed as the summation of the same orders of the power series of the 1$^{st}$ order bunching factors.

For the photon distribution determined by the diagonal term, we have

$$\tilde{\rho}_{ph}^{N_e}(n_{N_e}, n_{N_e}) = \sum_{r_i = \{-1, 0, 1\}}^{\{i=1,2,\ldots N_e\}} \left( \prod_{i=1}^{N_e} \tilde{b}_i^{(r_i)} \right) \sum_{n_i}^{\{i=1,\ldots,N_e-1\}} \delta\left( \sum_{j=1}^{N_e} r_j \right) \prod_{i=2}^{N_e} M_{n_i, n_{i-1}} \\ \times M_{n_1, 0} M_{0, n_1 + \sum_{j=2}^{N_e} r_j}^\dagger \prod_{i=1}^{N_e-2} M_{n_i + \sum_{j=i+1}^{N_e} r_j, n_{i+1} + \sum_{j=i+2}^{N_e} r_j}^\dagger M_{n_{N_e-1} + r_{N_e}, n_{N_e}}^\dagger$$ (S51)

According to the delta function $\delta\left(\sum_{j=1}^{N_e} r_j\right)$ shown in Eq. (S51), the orders of bunching factors of all electrons satisfy: $\sum_{j=1}^{N_e} r_j = 0$. Since only the 1st order bunching factor is involved and $\tilde{b}_i^{(0)} = 1$, the photon distribution could be expanded into the power series of $\tilde{b}_i^{(1)} \tilde{b}_j^{(-1)}$:

$$\rho_{ph}^{N_e}(n,n) = c_{0,n}^{(0)} + \sum_{i \neq j}^{N_e} \tilde{b}_i \tilde{b}_j^* c_{0,ij}^{(2)} + \sum_{i \neq j \neq k \neq l}^{N_e} \tilde{b}_i \tilde{b}_j^* \tilde{b}_k \tilde{b}_l^* c_{0,ijkl}^{(4)} + \sum_{\substack{i \neq j \neq k \neq \\ l \neq m \neq q}}^{N_e} \tilde{b}_i \tilde{b}_j^* \tilde{b}_k \tilde{b}_l^* \tilde{b}_m \tilde{b}_q^* c_{0,ijklmq}^{(6)} \cdots \quad (S52)$$

where we denote $\tilde{b}_i \equiv \tilde{b}_i^{(1)}$ and $\tilde{b}_i^* \equiv \tilde{b}_i^{(-1)}$ for simplicity. The coefficient $c^{2m}$ denotes the contribution form the 2m electrons within $N_e$ electrons, which is determined by the series $\{r_1, r_2, \ldots, r_j, \ldots, r_{N_e}\}$. For the diagonal terms, the number of terms $r_i = 1$ equals to the number of terms $r_i = -1$, due to $\sum_{j=1}^{N_e} r_j = 0$.

When $m = 0$, the contributions from all the electrons are $b^{(0)} = 1$, or say $\{r_i = 0, i = 1,2,\ldots,N_e\}$. Hence,

$$c_{0,n}^{(0)} = \sum_{n_i}^{\{i=0,1,\ldots,N_e-1\}} \delta_{n_0,0} \prod_{i=1}^{N_e} M_{n_i,n_{i-1}} M_{n_{i-1},n_i}^\dagger = \frac{1}{N_e|g|^2 + 1} \left(\frac{N_e|g|^2}{N_e|g|^2 + 1}\right)^n$$

which is a Poissonian distribution.

When $m = 1$, there are only two terms in the series $\{r_1, r_2, \ldots, r_j, \ldots, r_{N_e}\}$ are non-zero, where one term is 1 and the left term is -1. For a series with $r_i = 1$ and $r_j = -1$, if we assume $i < j$, it looks like

$$\{r_1 = 0, r_2 = 0, \ldots, r_{i-1} = 0, r_i = 1, r_{i+1} = 0, \ldots, r_{j-1} = 0, r_j = -1, r_{j+1} = 0, \ldots, r_{N_e} = 0\}$$

which leads to the bilinear term $b_i b_j^*$, the corresponding coefficient is

$$c_{ij}^{(2)}(N_e) = \left(\prod_{i=2}^{N_e} M_{n_i,n_{i-1}} M_{n_1,0}\right) M_{0,n_1}^\dagger M_{n_1,n_2}^\dagger \cdots M_{n_{i-1},n_i-1}^\dagger M_{n_i-1,n_{i+1}-1}^\dagger \cdots$$
$$\times \cdots M_{n_{j-1}-1,n_j}^\dagger M_{n_j,n_{j+1}}^\dagger \cdots M_{n_{N_e-1},n_{N_e}}^\dagger$$

With the help of numerical calculation, we find the positions of $r_i = 1$ and $r_j = -1$ do not affect the number of the coefficient shown above. Hence, the contribution to the final state $\rho_{ph}^{N_e}(n,n)$ from the bilinear terms can be written as

$$\sum_{i \neq j}^{N_e} \tilde{b}_i \tilde{b}_j^* c_{0,ij}^{(2)}(N_e) = c_{0,n}^{(2)} \sum_{i \neq j}^{N_e} \tilde{b}_i \tilde{b}_j^*$$

where $c_{0,n}^{(2)} = c_{0,ij}^{(2)}$ for arbitrary $i,j \in N_e$ satisfying $i \neq j$.

When $m = 2$, there are four terms in the series $\{r_1, r_2, \ldots, r_{N_e}\}$ are non-zero, where two terms are 1 and the other two terms are -1. The general form of the coefficient $c_{0,ijkl}^{(4)}$ in Eq. (S52) is still derived from Eq. (S51). For an arbitrary series $\{r_1, r_2, \ldots, r_i, \ldots r_j, \ldots, r_k, \ldots, r_l, \ldots, r_{N_e}\}$ with $r_i = r_j = 1$ and $r_k = r_l = -1$, (assuming $i < j < k < l$), satisfying $\sum_i r_i = 0$,

$$c_{0,ijkl}^{(4)}(N_e) = \left(\prod_{i=2}^{N_e} M_{n_i,n_{i-1}} M_{n_1,0}\right) M_{0,n_1}^\dagger M_{n_1,n_2}^\dagger \ldots M_{n_{i-1},n_i-1}^\dagger M_{n_i-1,n_{i+1}-1}^\dagger \ldots$$
$$M_{n_{j-2},n_{j-1}-1}^\dagger M_{n_{j-1}-1,n_j-2}^\dagger M_{n_j-2,n_{j+1}-2}^\dagger \ldots M_{n_{k-2}-2,n_{k-1}-2}^\dagger M_{n_{k-1}-2,n_k-1}^\dagger M_{n_k-1,n_{k+1}-1}^\dagger \ldots$$
$$M_{n_{l-2}-1,n_{l-1}-1}^\dagger M_{n_{l-1}-1,n_l}^\dagger M_{n_l,n_{l+1}}^\dagger \ldots M_{n_{N_e-1},n_{N_e}}^\dagger$$

Again, we found the sequence of the non-zero terms' index will not affect this coefficient, which leads to

$$\sum_{i \neq j \neq k \neq l}^{N_e} \tilde{b}_i \tilde{b}_j^* \tilde{b}_k \tilde{b}_l^* c_{0,ijkl}^{(4)}(N_e) = c_{0,n}^{(4)} \sum_{i \neq j \neq k \neq l}^{N_e} \tilde{b}_i \tilde{b}_j^* \tilde{b}_k \tilde{b}_l^*$$

The higher orders follow the similar properties.

For the off-diagonal terms, like the 1$^{st}$-order diagonal array, the summation of the bunching factors' index equals to -1 according to the term $\delta\left(\sum_{j=1}^{N_e} r_j + 1\right)$ in Eq. (S50). Hence, the 1$^{st}$-order diagonal array can be written as

$$\rho_{ph}^{N_e}(n, n+1) = \sum_i^{N_e} c_{1,i}^{(1)} \tilde{b}_i^* + \sum_{i \neq j \neq k}^{N_e} \tilde{b}_i \tilde{b}_j^* \tilde{b}_k^* c_{1,ijk}^{(3)} + \sum_{i \neq j \neq k \neq l \neq m}^{N_e} \tilde{b}_i \tilde{b}_j^* \tilde{b}_k \tilde{b}_l^* \tilde{b}_m^* c_{1,ijklm}^{(5)} + \cdots$$

where all the terms are composed of odd products of bunching factors, and the factors $c_1^{2m+1}$ represents the contribution from m $b^{(1)}$ and (m+1) $b^{(-1)}$. Similar to the diagonal terms, the coefficient is not relevant to the sequence of involved the electron index, hence we obtain

$$\rho_{ph}^{N_e}(n, n+1) = c_{1,n}^{(1)} \sum_i^{N_e} \tilde{b}_i^* + c_{1,n}^{(3)} \sum_{i \neq j \neq k}^{N_e} \tilde{b}_i \tilde{b}_j^* \tilde{b}_k^* + c_{1,n}^{(5)} \sum_{i \neq j \neq k \neq l \neq m}^{N_e} \tilde{b}_i \tilde{b}_j^* \tilde{b}_k \tilde{b}_l^* \tilde{b}_m^* + \cdots \quad (S53)$$

And

$$c_{1,n}^{(1)} = c_{1,i}^{(1)}(N_e) = \left(\prod_{i=2}^{N_e} M_{n_i,n_{i-1}} M_{n_1,0}\right) M_{0,n_1-1}^\dagger M_{n_1-1,n_2-1}^\dagger \ldots M_{n_{i-1},-1\, n_i}^\dagger M_{n_i,n_{i+1}}^\dagger \ldots M_{n_{N_e-1},n_{N_e}}^\dagger$$

$$c_{1,n}^{(3)} = c_{1,ijk}^{(1)}(N_e) = \left( \prod_{i=2}^{N_e} M_{n_i,n_{i-1}} M_{n_1,0} \right) M_{0,n_1-1}^{\dagger} M_{n_1-1,n_2-1}^{\dagger} \cdots M_{n_{i-1},-1 \, n_i-2}^{\dagger} M_{n_i-2,n_{i+1}-2}^{\dagger} \cdots$$

$$M_{n_{j-2}-2,n_{j-1}-2}^{\dagger} M_{n_{j-1}-2,n_j-1}^{\dagger} M_{n_j-1,n_{j+1}-1}^{\dagger} \cdots M_{n_{k-2}-1,n_{k-1}-1}^{\dagger} M_{n_{k-1}-1,n_k}^{\dagger} M_{n_k,n_{k+1}}^{\dagger} \cdots$$

$$M_{n_{N_e-1},n_{N_e}}^{\dagger}$$

| | |
|---|---|
| $\sum_{n=0}^{+\infty} c_{0,n}^{(0)} = 1,$ | $\sum_{n=0}^{+\infty} n c_{0,n}^{(0)} = g^2 N_e$ |
| $\sum_{n=0}^{+\infty} c_{0,n}^{(2)} = 0,$ | $\sum_{n=0}^{+\infty} n c_{0,n}^{(2)} = g^2$ |
| $\sum_{n=0}^{+\infty} c_{0,n}^{(2m)} = 0,$ | $\sum_{n=0}^{+\infty} n c_{0,n}^{(2m)} = 0 , for \, m \geq 2$ |

Table 1. the properties of the coefficients in the diagonal terms of the emission's density matrix.

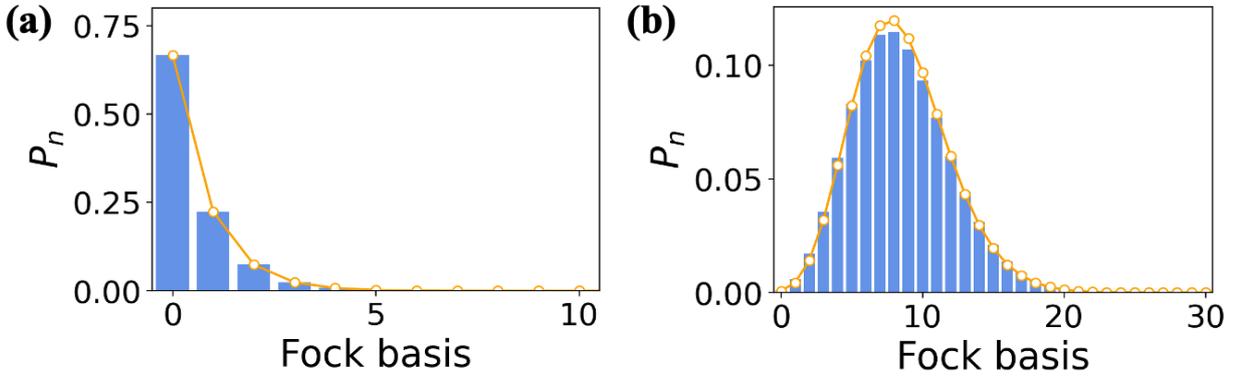

Figure S5: photon distribution for the (a) non-correlated QEWs and (b) modulation-correlated QEWs. The orange dotted line is simulation result, and the blue bar is calculated through analytical formula Eq. (S36).

## 7. Numerical computation method

In order to perform numerical calculation, we need the matrix form of the Hamiltonian, which is easy for the free energy of both electron and photon (Eq. (2)). Thus, the interaction Hamiltonian is

$$\hat{H}_I = -\frac{eA_0}{2\gamma m_e}\left[\hat{p}\cdot f(z)\left(\hat{a}e^{iq_z z}+\hat{a}^\dagger e^{-iq_z z}\right) + f(z)\left(\hat{a}e^{iq_z z}+\hat{a}^\dagger e^{-iq_z z}\right)\cdot\hat{p}\right]$$
$$= -\frac{eA_0}{2\gamma m_e}\left[\hat{a}\left(\hat{p}\cdot f(z)e^{iq_z z} + f(z)e^{iq_z z}\cdot\hat{p}\right) + \hat{a}^\dagger\left(\hat{p}\cdot f(z)e^{-iq_z z} + f(z)e^{-iq_z z}\cdot\hat{p}\right)\right]$$

If we define $h_I(z) = \left(\hat{p}\cdot f(z)e^{iq_z z} + f(z)e^{iq_z z}\cdot\hat{p}\right)$, then the interaction Hamiltonian could be expressed in the product space $|p\rangle\otimes|n\rangle$

$$\hat{H}_I = -\frac{eA_0}{2\gamma m_e}[h_I(z)\otimes\hat{a} + h_I^*(z)\otimes\hat{a}^\dagger] \tag{S54}$$

The coordinate-dependent function $h_I(z)$ in the wavevector basis becomes,

$$h_I(z) = \hbar\left(\left(\hat{k}+k_0\right)\cdot f(z)e^{iq_z z} + f(z)e^{iq_z z}\cdot\left(\hat{k}+k_0\right)\right)$$
$$= 2\hbar k_0 f(z)e^{iq_z z} + \hbar\left(\hat{k}\cdot f(z)e^{iq_z z} + f(z)e^{iq_z z}\cdot\hat{k}\right)$$

When projecting to the wavevector basis $\{|k\rangle\}$, the matrix electron is $M_{ij}(k) = \langle k_i|h_I(z)|k_j\rangle$. According to the conjugate relation $\langle z|k\rangle = e^{-ikz}$,

The matrix form of $e^{iq_z z}$ in wavevector basis is

$$T_{ij}(k) = \langle k_i|e^{iq_z z}|k_j\rangle = \sum_m e^{i(k_i-k_j+q_z)z_m}$$

For the Fourier transformation defined as $\hat{F} = \sum_{nm}|k_m\rangle\langle z_n|$, the matrix $T = \hat{F}e^{iq_z z}\hat{F}^\dagger$. And

$$\langle k_i|f(z)e^{iq_z z}|k_j\rangle = \sum_m f(z)e^{i(k_i-k_j+q_z)z}$$
$$\langle k_i|\hat{k}\cdot f(z)e^{iq_z z}|k_j\rangle = k_i\langle k_i|f(z)e^{iq_z z}|k_j\rangle$$
$$\langle k_i|f(z)e^{iq_z z}\cdot\hat{k}|k_j\rangle = k_j\langle k_i|f(z)e^{iq_z z}|k_j\rangle$$

In our assumption, the distribution function $f(z)$ is infinitely close to a rectangular function, then

$$\sum_m f(z)e^{i(k_i-k_j+q_z)z} \simeq \frac{2\sin(k_i-k_j+q_z)}{k_i-k_j+q_z}$$

Then we have the matrix form of interaction Hamiltonian. the simulation could be done by

$$|\psi_f\rangle = e^{-i(H_0+H_I)t/\hbar}|\psi_i\rangle$$

# 8. Classical and QED formulation of spontaneous and superradiant emission by a bunched electron beam

The QED derivation of spontaneous and superradiant radiation emission by modulation-correlated QEWs presented in the main text, is expected to be consistent with classical derivation of spontaneous emission by a random point-particles beam and superradiant emission by a bunched electron beam respectively. Here we compare the results for a model of electrons interaction with a dielectric waveguide.

**Classical expressions for spontaneous and superradiant emission**

We use the formulation derived in (34) for excitation of radiation modes by free point particles classical electrons. The classical field can be expanded in terms of a complete set of traveling radiation modes:

$$\begin{Bmatrix} E(r,\omega) \\ H(r,\omega) \end{Bmatrix} = \sum_q c_q(z,\omega) \begin{Bmatrix} \tilde{E}_q(r) \\ \tilde{H}_q(r) \end{Bmatrix} \tag{S55}$$

where $\{\tilde{E}_q(r), \tilde{H}_q(r)\} = \{\tilde{\mathcal{E}}_q(r_\perp), \tilde{\mathcal{H}}_q(r_\perp)\} e^{ik_{qz}z}$ are the modes of the longitudinally uniform interaction volume in the absence of interaction, and $c_q(z,\omega)$ is the slow varying amplitude of transverse mode q due to interaction with a classical current:

$$J(r,\omega) = \int_{-\infty}^{\infty} J(r,t) e^{i\omega t} dt \tag{S56}$$

From Maxwell equations, the following expression is derived for the development of each mode in an axially uniform structure or a waveguide [ ]:

$$\frac{d\tilde{C}_q(z,\omega)}{dz} = -\frac{1}{4\mathcal{P}_q} \int d^2 r_\perp \, \tilde{J}(r,\omega) \cdot \tilde{\mathcal{E}}_q^*(r_\perp) \, e^{-ik_{qz}z} \tag{S57}$$

where $\mathcal{P}_q = \frac{1}{2} \operatorname{Re}\{\int\int (\tilde{\mathcal{E}}_q \times \tilde{\mathcal{H}}_q^*) \cdot \hat{e}_z \, d^2 r_\perp\} = \frac{|\tilde{\mathcal{E}}_q(r_\perp=0)|^2}{2Z_q} A_{em,q}$ is the normalization power of the mode, $Z_q$ is the mode impedance and $A_{em,q}$ is the effective cross-sectional area of the mode.

For a single point particle electron, we substitute $J(r,t) = -ev(t)\delta(r - r_e(t))$. Integration of (S57) along the interaction region with $c_q(0,\omega) = 0$ (for spontaneous emission), results in:

$$C_q^{out} = -\frac{1}{4\mathcal{P}_q} \Delta W_{qe} \tag{S58}$$

$$\Delta W_{qe}^{(0)} = -e \int_{-\infty}^{+\infty} v_e^{(0)}(t) \cdot \tilde{E}_q^* \left(r_e^{(0)}(t)\right) e^{i\omega t} dt$$

For a beam of electrons:

$$C_q(\omega) = -\frac{1}{4\mathcal{P}_q} \sum_j \Delta W_{qj} \tag{S59}$$

$$\Delta W_{qj} = -e \int_{-\infty}^{+\infty} v_j(t) \cdot \tilde{E}_q^* \left(r_j(t)\right) e^{i\omega t} dt$$

The electrons velocity and trajectory in the absence of interaction with radiation field (no self-interaction) are

$$v_j(t) = v_j^{(0)}(t), \qquad r_j(t) = r_j^{(0)}(t) \tag{S60}$$

Then to 0 order (no self-interaction with the radiation field):

$$\Delta W_{qi}^{(0)} = -e \int_{-\infty}^{+\infty} v_j^{(0)}(t) \cdot \tilde{E}_q^* \left(r_j^{(0)}(t)\right) e^{i\omega t} dt \tag{S61}$$

From Parceval theorem,

$$\frac{dW_q}{d\omega} = \frac{2}{\pi} P_q |C_q|^2 = \frac{1}{8\pi P_q} \left| \sum_{j=1}^{N_e} \Delta W_{qj} \right|^2 \quad \text{(S62)}$$

Assume all electrons pass the same trajectories and have the same velocities, but enter the interaction region at different phases:

$$\Delta W_{qj} = \Delta W_{qe}^{(0)} e^{i\omega t_{0j}} \quad \text{(S63)}$$

Thus, the spectral radiation per transverse mode q is:

$$\overline{\frac{dW_q}{d\omega}} = \frac{\left|\Delta W_{qe}^{(0)}\right|^2}{8\pi P_q} \overline{\left| \sum_{j=1}^{N_e} e^{i\omega t_{0j}} \right|^2} \quad \text{(S64)}$$

We define the bunching coefficient as

$$M_b(\omega) = \frac{1}{N_e} \sum_{j=1}^{N_e} e^{i\omega t_{0j}} \quad \text{(S65)}$$

Then:

$$\overline{\frac{dW_q}{d\omega}} = N_e^2 \, \overline{|M_b(\omega)|^2} \, \frac{\left|\Delta W_{qe}^{(0)}\right|^2}{8\pi P_q} \quad \text{(S66)}$$

For random electron entrance times $t_{0j}$

$$\overline{\left| \sum_{j=1}^{N_e} e^{i\omega t_{0j}} \right|^2} = \sum_{j=1}^{N_e} 1 + \overline{\sum_{j\neq k} e^{i\omega(t_{0j}-t_{0k})}} = N_e \quad \text{(S67)}$$

Thus, the bunching factor is $\overline{|M_b(\omega)|^2} = \frac{1}{N_e}$, and the classical spontaneous emission is:

$$\left(\frac{dW_q}{d\omega}\right)_{sp} = N_e \frac{\left|\Delta W_{qe}^{(0)}\right|^2}{8\pi P_q} = N_e \left(\frac{dW_q}{d\omega}\right)_{sp,e} \quad \text{(S68)}$$

If all electrons are bunched in a pulse shorter than the period of the radiation frequency:

$$\omega t_{0j} \ll \pi \;\; \forall j$$

or are bunched in a periodic (or sparsely periodic) distribution:

$$|\omega t_{0j} - m_j 2\pi| \ll \pi \;\; \forall j$$

where $m_j$ could be any integer, then

$$\sum_{j=1}^{N_e} e^{i\omega t_{0j}} = N_e \quad \text{(S69)}$$

The bunching coefficient is $M_b(\omega) = 1$, which gives the expression for superradiant emission:

$$\left(\frac{dW_q}{d\omega}\right)_{sr} = N_e^2 \frac{\left|\Delta W_{qe}^{(0)}\right|^2}{8\pi P_q} = N_e^2 \left(\frac{dW_q}{d\omega}\right)_{sp,e} \quad \text{(S70)}$$

In general, the radiation from a bunch of electrons can be written as

$$\left(\frac{dW_q}{d\omega}\right)_{sr} = N_e^2 |M_b|^2 \left(\frac{dW_q}{d\omega}\right)_{sp,e} \tag{S71}$$

where the bunching coefficient $|M_b|^2$ satisfies

$$\frac{1}{N_e} \leq |M_b|^2 \leq 1 \tag{S72}$$

The analysis, so far, is valid for any kind of interaction mechanism, such as undulator radiation (FEL) and Cerenkov radiation. Let us specify the case of a dielectric waveguide.

$$v_e^{(0)}(t) = v \cdot \hat{e}_z$$
$$t = z/v$$

$$\tilde{\boldsymbol{E}}_q^*\left(\boldsymbol{r}_e^{(0)}(t)\right) = \hat{e}_z \mathcal{E}_{qz}(\boldsymbol{r}_{e\perp}) e^{iq_z z} + \hat{e}_\perp \mathcal{E}_{q\perp}(\boldsymbol{r}_{e\perp}) e^{iq_z z}$$

We get

$$\Delta W_{qe}^{(0)} = -e \int_0^L \mathcal{E}_{qz}(\boldsymbol{r}_{e\perp}) \cdot e^{i(\frac{\omega}{v} - q_z)z} dz \tag{S73}$$

Under the synchronism condition $\frac{\omega}{v} - q_z = 0$, and then $\Delta W_{qe}^{(0)} = -e\mathcal{E}_{qz}L$, where $L$ is the length of the interaction region. The mode power normalization factor is $P_q = \frac{1}{2Z_q}|\mathcal{E}_{q\perp}(r_{\perp 0})|^2 A_{em,q}$.

Thus, the spontaneous spectral radiation emergy per mode per electron is

$$\left(\frac{dW_q}{d\omega}\right)_{sp,e} = \frac{1}{8\pi P_q} \left|\Delta W_{qe}^{(0)}\right|^2 = \frac{e^2 \mathcal{E}_{qz}^2 L^2}{8\pi |\mathcal{E}_{q\perp}|^2 A_{em}} 2Z_q = \frac{e^2 \eta_q^2 L^2}{4\pi A_{em}} Z_q \tag{S74}$$

where $\eta_q = |\mathcal{E}_{qz}|/\mathcal{E}_{q\perp}$, and $A_{em}$ is the effective cross-sectional area of the cavity waveguide.

To compare to the quantum result, we need to calculate the radiative emission energy per cavity mode. Since the analysis so far is for a single transverse waveguide mode $q$, we need to multiply the spectral energy by the spacing of the longitudinal mode of the resonator:

$$\Delta\omega = \frac{2\pi c}{L_c n_{eff}}$$

where $n_{eff} = k_{qz}/k$, $k = \omega/c$ and $L_c$ is the circumference of the resonator. Thus, the radiation energy is

$$\Delta W_{sp,mode} = \left(\frac{dW_q}{d\omega}\right)_{sp,e} \cdot \frac{2\pi c}{L_c n_{eff}} = \frac{e^2 \eta_q^2 L^2}{2L_c A_{em}} \frac{cZ_q}{n_{eff}}$$
$$= \frac{\eta_q^2}{2n_{eff}^2} \frac{e^2 L^2}{\epsilon_0 V_{mode}} \tag{S75}$$

where $V_{mode} = A_{em} \cdot L_c$ and $Z_q = \sqrt{\mu_0/\epsilon_0}/n_{eff}$.

The average photon number of spontaneous emission per electron per cavity mode is:

$$n_{sp,mode}^{(class)} = \frac{\Delta W_{sp,mode}}{\hbar\omega} = \frac{\eta_q^2}{2n_{eff}^2} \frac{e^2 L^2}{\epsilon_0 V_{mode}} \frac{1}{\hbar\omega} \tag{S76}$$

The average photon number of spontaneous photons emitted by $N_e$ uncorrelated electrons per cavity mode is:

$$n_{sp,mode}^{(N_e)} = N_e n_{sp,mode}^{(class)} \tag{S77}$$

The corresponding superradiance emission per mode for $N_e$ electrons is

$$n_{sr,mode}^{(N_e)} = M_b^2 N_e^2 n_{sp,mode}^{(class)} \tag{S78}$$

**QED expressions for spontaneous and superradiant emission**

In the QED analysis of the main text, we found for the spontaneous emission of $N_e$ QEWs into a single cavity mode (Eq. 17):

$$\left\langle n_{ph}^{(N_e)} \right\rangle = N_e |g|^2 \tag{S79}$$

For modulation correlated QEWs beam (superradiant emission) (Eq. 20A):

$$\langle n_{ph} \rangle = N_e |g|^2 + N_e(N_e - 1)|g|^2 |b^{(1)}|^2 \approx |g|^2 |b^{(1)}|^2 N_e^2 \tag{S80}$$

where the last approximate equality is valid for $N_e \gg 1$.

There is difference between the bunching coefficient $b^{(1)}$ of a single modulated QEW in the QED model and the bunching factor $M_b$ that is defined in the classical formulation on an ensemble of electrons. Yet comparison of Eqs. S77 and S79 shows almost the same functional scaling of spontaneous and superradiant emissions, with $M_b$ corresponding to $b^{(1)}$ and $n_{sp,mode}^{(class)}$ corresponding to $|g|^2$.

To complete the correspondence, we derive the explicit expression of the QED coupling coefficient in the case of a single cavity mode Eq. (S9):

$$g = i\eta_q \frac{eA_0 L}{\hbar}$$

$$A_0 = \sqrt{\frac{\hbar}{2n_{eff}^2 \epsilon_0 \omega_c V_c}} \tag{S81}$$

$$|g|^2 = \frac{\eta_q^2}{2n_{eff}^2} \frac{e^2 L^2}{\epsilon_0 V_c} \frac{1}{\hbar \omega_c}$$

which is identical with Eq. (S76).